\documentclass[aip,jcp,twocolumn,superscriptaddress,10pt]{revtex4-1}

\usepackage{color}
\usepackage{fancyhdr} 
\usepackage{varwidth} 
\usepackage{graphicx,subfigure}  
\usepackage{bm} 
\usepackage{listings}
\usepackage{enumerate}%
\usepackage[colorlinks,citecolor=blue,linkcolor=blue]{hyperref}
\usepackage{amsmath}
\usepackage{threeparttable}
\usepackage{graphicx}
\usepackage{amssymb}
\usepackage{amsthm}
\usepackage{ulem}
\usepackage{rotating}
\usepackage{multirow}

\makeatletter
\def\@email#1#2{%
 \endgroup
 \patchcmd{\titleblock@produce}
  {\frontmatter@RRAPformat}
  {\frontmatter@RRAPformat{\produce@RRAP{*#1\href{mailto:#2}{#2}}}\frontmatter@RRAPformat}
  {}{}
}%
\makeatother

\begin{document}
\title{A quantitative theory and atomistic simulation study on the soft-sphere crystal-melt interfacial properties: II. Interfacial free energies}

\author{Ya-Shen Wang}
\affiliation{State Key Laboratory of Precision Spectroscopy, School of Physics and Electronic Science, East China Normal University, Shanghai 200241, China}
\author{Zun Liang}
\affiliation{State Key Laboratory of Precision Spectroscopy, School of Physics and Electronic Science, East China Normal University, Shanghai 200241, China}
\author{Brian B. Laird}
\affiliation{Department of Chemistry, University of Kansas, Lawrence, KS 66045, USA}
\affiliation{Freiburg Institute for Advanced Studies (FRIAS), Albert-Ludwigs-Universit\"{a}t Freiburg, Albertstra{\ss}e 19, 79104 Freiburg im Breisgau, Germany}
\author{Yang Yang}
\affiliation{State Key Laboratory of Precision Spectroscopy, School of Physics and Electronic Science, East China Normal University, Shanghai 200241, China}

\begin{abstract}
This study proposes a new method for predicting the crystal-melt interfacial free energy ($\gamma$) using the Ginzburg-Landau (GL) model, enhanced by atomistic simulation data for more accurate density wave profiles. The analysis focuses on the soft-sphere system governed by an inverse power potential that stabilizes both BCC and FCC phases. Equilibrium molecular dynamics (MD) simulations are used to obtain density wave amplitude distributions, which serve as inputs for the GL model to predict $\gamma$ and its anisotropy.  The predicted $\gamma$ values exhibit strong agreement with prior benchmark simulation experimental studies, particularly for FCC crystal-melt interfaces (CMIs). The GL models for the CMI $\gamma$ are proved to be both computationally efficient and reasonably valid, offering quantitative predictions of $\gamma$ while providing insights into the factors controlling its magnitude and anisotropy. Key improvement is suggested for the variational procedure used in the two-mode CMI free energy functionals, and potential upgrades to the GL model are also proposed to further enhance predictive accuracy.

\end{abstract}
\maketitle

\section{introduction}
The crystal-melt interfacial free energy, $\gamma$, is a fundamental parameter that shapes the morphological evolution of the solidification microstructure and governs the kinetics of crystal nucleation. Defined as the reversible work required to create a unit area of the crystal-melt interface  {(CMI)}, $\gamma$ determines the nucleation barrier in undercooled melts and drives transitions between planar, cellular, and dendritic growth modes. Its capillary anisotropy influences both the branching patterns and growth rates of dendritic solidification {.~\cite{Hoyt01,Asta09}} However, direct experimental measurement of $\gamma$ is challenging. (i) Anisotropy measurements are largely limited to a few transparent organic systems {.~\cite{Kauerauf98}} (ii) For metallic systems of practical interest, $\gamma$ is typically inferred indirectly from crystal nucleation rates via classical nucleation theory, where the accuracy is often compromised by the potential for heterogeneous nucleation due to impurities. In dendrite-forming systems with atomically rough interfaces, the anisotropy of $\gamma$ is weak, making experimental quantification especially difficult.

Advanced simulation and computational approaches, such as the cleaving method,~\cite{Davidchack00,Broughton86} capillary wave fluctuations,~\cite{Hoyt01} mold integration,~\cite{Espinosa14} metadynamics,~\cite{Angioletti-Uberti10} tethered Monte Carlo,~\cite{Fernandez12} and several other thermodynamic integration methods~\cite{Schilling09,Benjamin14,Bultmann20} have enabled reliable predictions of the magnitude and anisotropy of $\gamma$ for elemental and alloy CMIs. These computational insights have advanced understanding of dendritic microstructure evolution,~\cite{Asta09} crystal polymorphism,~\cite{Schoonen22} and interfacial metastable phase transition, such as premelting.~\cite{Hoyt09,Yang13} The simulation and computational studies have notable trends in $\gamma$ with respect to crystal structure and CMI orientation.~\cite{Sun04,Sun04b,Davidchack05} Key observations include: (i) BCC CMIs exhibit lower $\gamma$ values than FCC CMIs. (ii) A consistent ordering of $\gamma_{100} > \gamma_{110} > \gamma_{111}$ is observed across a range of FCC systems. The anisotropy for FCC CMIs  is generally a few percent, significantly larger than the $\sim$0.5\% anisotropy observed for BCC CMIs. Despite these computational successes, a fundamental theoretical framework that fully explains the crystal structure- and material-dependent factors controlling the magnitude and anisotropy of $\gamma$ is still lacking.

For materials with more realistic interatomic interactions, the application of density functional theory remains scarce.~\cite{Oxtoby82,Schoonen22b} More widely adopted theoretical models are Landau-Ginzburg-type approaches, such as the phase-field crystal (PFC) model~\cite{Elder07,Wu07,Majaniemi09,Toth14} and the Ginzburg-Landau (GL) model.~\cite{Shih87,Wu06,Wu16} Despite their computational efficiency, these models are often criticized for being too generic to accurately capture the detailed structure and density fields of CMIs. Consequently, they tend to exhibit limited accuracy in predicting the magnitude and anisotropy of $\gamma$. The GL and PFC models have been used to explore the relationship between anisotropy and crystal symmetry. They have been applied to both BCC and FCC CMIs to investigate crystalline anisotropies.~\cite{Shih87,Wu06,Wu07,Majaniemi09,Greenwood10,Toth14,Wu16} 
However, theoretical predictions of the CMI $\gamma$ from these models have not shown good agreement with computer experiments. For FCC CMIs, such as hard-sphere models~\cite{Oettel12} and metals,~\cite{Nourian-Avval18,Wu16} both models significantly underestimated the magnitude of $\gamma$ (by up to an order of magnitude) and failed to capture the correct anisotropy sequence across crystalline orientations. In BCC CMIs, while the models produced reasonable estimates of the weak anisotropy in $\gamma$,~\cite{Wu06,Wu07,Jaatinen09} they still exhibited varying degrees of underestimation and overestimation of its magnitude.

The limited accuracy of the GL and PFC models in predicting CMI $\gamma$ and its anisotropy has raised skepticism about their reliability for quantitative predictions of $\gamma$'s dependence on crystal structure and material properties, despite their computational simplicity. As noted by Oettel \textit{et al}.,~\cite{Oettel12} while the GL and PFC models can predict certain bulk phase properties, they perform poorly in estimating interfacial properties. This limitation may stem from the fact that the density wave profiles across the interface, derived via global optimization of the free energy functional with respect to density wave amplitudes, do not account for the local atomistic packing structure at the  {CMI.}~\cite{Wu06,Wu07} We hypothesize that this omission underlies the models' inaccurate predictions of $\gamma$ values and anisotropy. To date, few studies have attempted to incorporate density wave profiles based on local particle packing at the  {CMI} into the GL or PFC frameworks to improve interfacial energy predictions.

This paper investigates a model system based on an inverse power potential (soft spheres) that stabilizes both BCC and FCC phases. Using equilibrium molecular dynamics simulations, we accurately measure the density field distribution of CMIs by tracking local atomistic trajectories. The resulting density wave amplitude distribution functions serve as input for predicting CMI $\gamma$ within the GL model framework. Our results show that the predicted values of CMI $\gamma$ and its anisotropy closely align with previous benchmark computational simulation results. To the best of our knowledge, this level of agreement surpasses all prior studies using the GL or PFC models to predict $\gamma$. Our findings challenge the conventional approach of using global optimization to derive the CMI density field in the two-mode GL model for FCC CMIs, suggesting it may be flawed. By validating the GL model's quantitative predictive capability, we also demonstrate its ability to provide physical insights into key variations in $\gamma$, including the primary factors that govern its magnitude and anisotropy.

 {\section{GL models of equilibrium BCC and FCC CMIs}}

The GL models employed in this study are based on the single-mode Ginzburg-Landau (GL) model for equilibrium BCC crystal-melt interfaces (CMIs) developed by Shih \textit{et al}.~\cite{Shih87} and Wu \textit{et al}.,~\cite{Wu06} as well as the two-mode GL model for equilibrium FCC CMIs developed by Wu \textit{et al}.~\cite{Wu16} In both models, the free energy functional $F$ of the inhomogeneous CMI system, formulated within the density-functional theory (DFT) framework at the crystal-melt coexistence temperature $T_\mathrm{m}$, is expressed as a functional of GL order parameters. These order parameters correspond to the amplitudes of density waves associated with specific wave vectors -- the principal set and the second set of reciprocal lattice vectors for FCC CMIs, and the principal reciprocal lattice vectors for BCC CMIs  { (see Eqs.(S1) and Eqs.(S2) in Supplemental Materials.~\cite{SM})}

In the formalism of these two GL models, the excess free energy is expanded as a power series of the GL order parameters. The coefficients of this series are specified using cDFT for inhomogeneous fluids with small density fluctuations  {(see Eqs.(S6)-(S7) and Eqs.(S10)-(S11) in Supplemental Materials.~\cite{SM})} By incorporating the distinct GL order parameters in the interfacial regions, the GL free energy functional can be reduced to specific expressions for the interfacial free energy $\gamma$ (defined as the excess free energy per unit area) for various crystalline orientations. These include FCC(100), FCC(110), FCC(111), BCC(100), BCC(110), and BCC(111),  {as shown in Eqs.(S12)-(S17) in Supplemental Materials.~\cite{SM}}

Further details on the theoretical formalism, including the density functionals, free energy functionals, and classifications of density wave amplitudes (GL order parameters), are provided in  {Supplemental Materials.~\cite{SM}}

\section{Computational Methods}

\subsection{Inverse-power repulsive potential}
\label{subsec-ss}

The present study employs the soft-sphere (SS) model system, which is described by the pairwise inverse-power repulsive potential:
\begin{equation}
U(r_{ij}) = \epsilon \left( \frac{\sigma}{r_{ij}} \right)^n,
\label{ss}
\end{equation}
where $r_{ij}$ is the particle separation, and $\sigma$ and $\epsilon$ set the length and energy scales, respectively. The repulsive range decreases with increasing $n$, approaching the hard-sphere (HS) limit as $n \to \infty$. The potential can alternatively be characterized by the softness parameter $s = 1/n$.

This potential series is a suitable candidate for comparing FCC and BCC CMI properties, as the free energy difference between the FCC and BCC phases is small for $n=6, 7, 8$, and metastable FCC or BCC crystal-melt interfaces remain stable even during long MD simulations.~\cite{Davidchack05} 
Our study focuses on FCC and BCC CMIs modeled using the SS system with $n = 6$. We adopt the interaction truncation method of Heyes \textit{et al}.,~\cite{Heyes05} with the cutoff distance defined as $r_c = \sigma [\epsilon / (10^{-4} k_{\mathrm{B}} T)]^{1/6}$. To ensure smooth decay of the potential and force at $r_c$, we employ the smoothing function proposed by Morris \textit{et al}.~\cite{Morris02}:
\begin{equation}
f(x)=\begin{cases}
1&x<0\\ 
1-3x^{2}+2x^{3}&0\leq x<1\\ 
0&1\leq x
\label{smoothcut}
\end{cases}
\end{equation}
here, $x = (r_{ij} - r_m) / (r_c - r_m)$, with $r_m = 0.95 r_c$. While the cutoff distance $r_c$ preserves scaling, the smoothing function could introduce non-scaling behavior. However, as shown in our recent work,~\cite{Wang24} within the temperature-pressure range studied, no significant non-scaling behavior is observed. For simulations and computations, we employ dimensionless reduced units, denoted by superscript asterisks. These units are defined by scaling particle mass, distances, energies, and temperatures using the units $m$, $\sigma$, $\epsilon$, and $\epsilon / k_{\mathrm{B}}$, respectively.

A key advantage of the SS model is that its partition function depends only on the quantity $\epsilon \sigma^n$, not separately on $\epsilon$ and $\sigma$.~\cite{Hansen76} As a result, all excess thermodynamic quantities depend solely on the dimensionless parameter
\begin{equation}
\Gamma_n=\rho\sigma^3(k_{\mathrm{B}}T/\epsilon)^{-3/n} = \rho^* (T^*)^{-3/n}
\end{equation}
where $\rho^* = \rho \sigma^3$ and $T^* = k_{\mathrm{B}} T / \epsilon$ are the reduced density and temperature, respectively. This scaling law implies that systems with equal $\Gamma_n$ exhibit identical thermodynamic behavior, as discussed in detail in our recent work.~\cite{Wang24} For crystal-melt interfaces (CMIs) along the coexistence boundary, the interfacial free energy $\gamma$ exhibits a power-law scaling with respect to the reduced melting temperature $T_{\mathrm{m}}^*$ and the repulsive power $n$ in the SS potential:
\begin{equation}
\gamma^{*}\equiv \gamma\sigma^2\epsilon^{-1}=\gamma_1T_{\mathrm m}^{*1+2/n}.
\label{eq.scaling.gamma}
\end{equation}
where $\gamma^*$ is the reduced interfacial energy, and $\gamma_1$ denotes the value of $\gamma^*$ at $T_{\mathrm{m}}^* = 1$.

\subsection{Simulation details}

MD simulations are conducted using the parallel MD simulation code LAMMPS.~\cite{Plimpton95} The current study employs constant $NVT$, $NpT$, and $NpH$ ensembles, utilizing the same thermostat and barostat methods as described in Refs.~\onlinecite{Wang22,Zhang22}. The thermostat and barostat relaxation times are set to 0.05 and 0.5, respectively, with an MD time step of $\delta t^* = 0.01$ to ensure accurate sampling of temperature and pressure relative to the imposed values.

Equilibrium MD simulations are used to compute the GL order parameter profiles (OPPs for short) of CMIs and to measure the static structure factors of bulk melt phases. All simulations are performed with periodic boundary conditions (PBC) to eliminate surface effects and ensure bulk-like behavior.

Crystal-melt coexisting states are carefully prepared using the equilibrium MD techniques of Du \textit{et al}.~\cite{Du18} and Liang \textit{et al}.~\cite{Liang21} The $z$-axis is defined as the direction normal to the CMI, while the $x$- and $y$-axes lie parallel to the CMIs. Six crystallographic orientations are examined in this study: (100), (110), and (111) for both FCC and BCC CMIs. Details on the simulation cell geometries and sizes for the various CMI systems are provided in  {Tab.~S1 of the Supplemental Materials.~\cite{SM}} For the FCC(100) CMI, six different coexistence conditions along the $p$-$T$ coexistence boundary are investigated, as reported in our recent study on the same $n = 6$ SS system.~\cite{Wang24}

\begin{figure}[!htb]
\centering
 \includegraphics[width=0.45\textwidth]{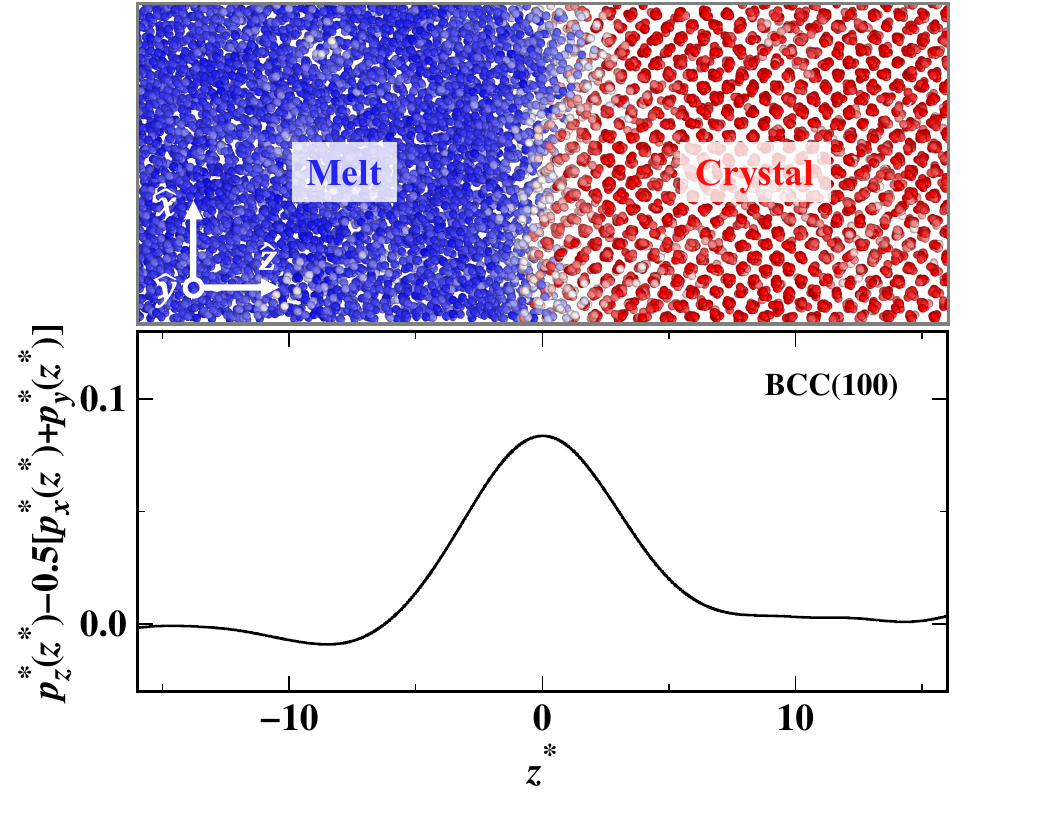}
 \caption{Snapshot of the equilibrium BCC(100) CMI (top panel), particles are color-coded based on the structural order parameter. Time-averaged stress profile (bottom panel), the stress is monitored to ensure that the hydrostatic condition is satisfied in the bulk phases during the setup of the equilibrium CMIs. The interface normal direction is denoted as $z^*$, where $z^* = 0$ corresponds to the mean position of the fluctuating CMI, $\langle \xi^*(t^*) \rangle$. Regions with $z^* < 0$ correspond to the melt, while $z^* > 0$ corresponds to the crystal. Superscript asterisks indicate dimensionless units.}
 \label{fig1}
 \end{figure}

To construct the simulation box containing coexisting crystal and melt phases, separate bulk crystal and melt samples are first prepared at the melting temperature and corresponding coexistence pressure, ensuring equal cross-sectional area ($A$) for both samples. The samples are then joined at their common cross-section and subjected to PBCs along the $z$-axis. The system is equilibrated using the $Np_zAH$ ensemble, with simulations run for $5 \times 10^6$ MD steps. The cross-sectional dimensions of the simulation cells are iteratively adjusted to account for thermal expansion of the crystal lattice, ensuring accurate coexistence conditions (temperature and pressure) are achieved. This adjustment process is repeated until precise coexistence is established. During equilibrium simulations, the stress within the crystal and melt phases is monitored to ensure that the hydrostatic condition is maintained.~\cite{Lu22}  {Fig.~\ref{fig1}} presents a representative snapshot from the equilibrated $NVT$ simulations of a BCC(100) CMI, along with the corresponding coarse-scale stress profiles.

To calculate the GL OPPs of the CMIs, we follow the initial $Np_zAH$ run with $2\times10^6$ steps of $NVT$ MD simulations. The $z$-dimension length ($L_z$) for the $NVT$ simulations is set to the average $L_z$ obtained from the equilibrium $Np_zAH$ runs. Initial configurations for the $NVT$ simulations are selected from the $Np_zAH$ trajectories, ensuring that the chosen $L_z$ matches the average value. The calculation of the GL OPPs based on MD simulation uses $4\times10^6$ instantaneous CMI configurations generated by the each time step during the $NVT$ runs (PBCs result in two independent interfaces within each simulation box).

To prevent artificial broadening of interfacial profiles caused by the Brownian motion of the crystal slab, we subtract the linear momentum of the innermost crystal layers, ensuring no significant drift in the positions of the CMI or the crystal slab during the $NVT$ production simulations. Particle coordinates from every MD step of the $NVT$ simulations are used to compute the equilibrium density fields and the GL OPPs.

To compute the static structure factors $S(k)$ of bulk liquids, we conduct independent simulations of a bulk melt containing 4,000 SS particles at the crystal-melt coexistence temperature and pressure. Each simulation runs for $5 \times 10^6$ MD steps, with 5000 trajectories used in the computation.

\begin{table*}[!htb]
\caption{Values of input parameters used in the GL models for CMI $\gamma$ in both BCC and FCC systems are provided. These include the magnitudes (in reduced units) of $|\vec{K}_{110}^{*}|$ for the twelve $[110]$ RLVs in BCC, as well as $|\vec{K}_{111}^{*}|$ and $|\vec{G}_{200}^{*}|$ for the eight $[111]$ and six $[200]$ RLVs in FCC. Also included are the static structure factors and the derivatives of the direct correlation function at these wavenumbers.}
\begin{ruledtabular}
\begin{tabular}{lccccccccccccc}
&
$T^{*}_\mathrm{m}$&
$p^{*}$&
$\rho^{*}_\mathrm{m}$&
$|\vec{K}^{*}_{110}|$&
$|\vec{K}^{*}_{111}|$&
$|\vec{G}^{*}_{200}|$&  
$S(|\vec{K}^{*}_{110}|)$&
$S(|\vec{K}^{*}_{111}|)$&
$S(|\vec{G}^{*}_{200}|)$&
$C'(|\vec{G}^{*}_{200}|)$&
$C''(|\vec{K}^{*}_{110}|)$&
$C''(|\vec{K}^{*}_{111}|)$&
$C''(|\vec{G}^{*}_{200}|)$\\
\hline
BCC &0.988&100&2.297&9.340&-&-&3.066&-&-&-&-0.989&-&-\\
FCC&0.995&103&2.319&-&9.111&10.521&-&2.960&1.180&-0.639&-&-1.305&-0.049\\
\end{tabular}
\end{ruledtabular}
\label{tab1}
\end{table*}

\subsection{Validation of the GL model using MD simulation data}

To validate the GL model for CMI $\gamma$ in both FCC and BCC CMIs, we compared the $\gamma$ values predicted by the analytical expressions  {[Eqs.(S12)-(S17) in Supplemental Materials~\cite{SM}]} with the precise values measured by Davidchack and Laird using the cleaving method based on atomistic simulations.~\cite{Davidchack05}

The values of the melting temperature $T_\mathrm{m}$, reciprocal wave vectors of the bulk crystals ($|\vec{K}^*_{110}|$ for BCC, $|\vec{K}^*_{111}|$ and $|\vec{K}^*_{200}|$ for FCC), and the density of the uniform bulk melt phase $\rho_\mathrm{m}$, measured from the equilibrium CMI simulation systems, are listed in Tab.~\ref{tab1}. The multiplicative coefficients in  {Eqs.(S12)-(S17) (provided in Tab.~S1~\cite{SM})} are determined from the density amplitudes ($u_{\mathrm{s}}$, $v_{\mathrm{s}}$) in the bulk crystal phase (as detailed in Subsection~\ref{subsec-OPPs}), the static structure factors at wave numbers $k^* = |\vec{K}^*_{110}|$, $|\vec{K}^*_{111}|$, or $|\vec{K}^*_{200}|$, and the derivatives of the direct correlation function [$C'(|\vec{G}^*_{200}|)$, $C''(|\vec{K}^*_{110}|)$, $C''(|\vec{K}^*_{111}|)$, $C''(|\vec{G}^*_{200}|)$], as listed in Tab.~\ref{tab1}.

The order parameter profiles (density wave amplitudes) along the $z$-axis are obtained by aligning each fluctuating CMI position [$\xi^*(t^*)$] and calculating time-spatial averages in the corresponding reference frame:
\begin{equation}
\begin{aligned}
u_i(z^*)&=\left|\langle\hat{u}_i[z^*-\xi^*(t^*),t^*]\rangle\right|,\\
v_i(z^*)&=\left|\langle\hat{v}_i[z^*-\xi^*(t^*),t^*]\rangle\right|,
\label{eq:uvopro}
\end{aligned}
\end{equation}
where the alignment procedure eliminates artificial broadening caused by the Brownian-like random motion of the CMI and the crystal.

The instantaneous density amplitudes are calculated from the Fourier transform of the instantaneous particle number density field $\rho(\vec{r}, t)$:
\begin{equation}
\begin{aligned}
\hat{u}_i(z,t)&=\frac{1}{A_{xy}\Delta^*_z}\int_0^{L_x}\int_0^{L_y}\int_{z-\frac{\Delta^*_z}{2}}^{{z+\frac{\Delta^*_z}{2}}}\mathrm{d}x\mathrm{d}y\mathrm{d}z \rho(\vec{r},t)\mathrm{\bf{e}}^{i \vec{K}_{i} \cdot \vec{r}} \\
\hat{v}_i(z,t)&=\frac{1}{A_{xy}\Delta^*_z}\int_0^{L_x}\int_0^{L_y}\int_{z-\frac{\Delta^*_z}{2}}^{{z+\frac{\Delta^*_z}{2}}}\mathrm{d}x\mathrm{d}y\mathrm{d}z \rho(\vec{r},t)\mathrm{\bf{e}}^{i \vec{G}_{i} \cdot \vec{r}},
\label{eq:2uvopro}
\end{aligned}
\end{equation}
where $A_{xy} = L_x L_y$ is the cross-sectional area, $\Delta^*_z=0.01$ is the bin size.

The order parameter profiles are computed for distinct categories of nonequivalent reciprocal-lattice vectors (RLVs), as listed in  {Tab.~S2 in the Supplemental Materials.~\cite{SM}} Statistical errors are estimated using around 1000 data blocks, each containing approximately 5000 successive MD steps.

Previous theoretical studies~\cite{Shih87, Wu06, Wu07, Wu16} have numerically solved the GL order parameter profiles (OPPs) for CMIs by minimizing the free-energy functional with respect to the order parameter profiles, subject to specific boundary conditions. In this study, we demonstrate that GL OPPs, calculated from well-equilibrated crystal-melt two-phase coexistence MD simulations with sufficient sampling, differ from those obtained via analytical minimization. Our results highlight the necessity of incorporating local atomistic packing information into the GL OPPs to achieve accurate predictions of the CMI free energy ($\gamma$) within the GL model framework.

With the multiplicative coefficients (Tab.~\ref{tab1}) and the GL order parameter profiles (OPPs) prepared, the CMI free energy ($\gamma$) is predicted by integrating  {Eqs.(S12)-(S17)~\cite{SM}} over the corresponding CMI region. The integration is performed using the Simpson's rule, ensuring high numerical accuracy. The systematic error from the numerical integration is found to be smaller than the statistical error, confirming the robustness of the method.

\section{Results and Discussion}

\subsection{Density and density wave amplitude profiles}
\label{subsec-OPPs}

\begin{figure*}[!htb]
\centering
\includegraphics[width=.99\textwidth]{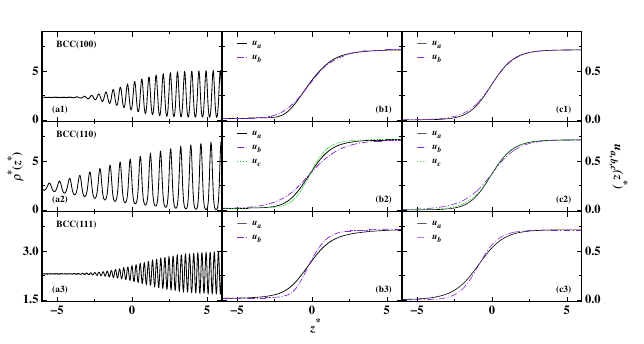}
\caption{(a1)-(a3) Fine-grained density profiles for the equilibrium BCC(100), BCC(110), and BCC(111) CMIs. (b1)-(b3) GL OPPs for the BCC(100), BCC(110), and BCC(111) CMIs, computed from equilibrium MD simulations using Eqs.(\ref{eq:uvopro}) and (\ref{eq:2uvopro}), with the $\hat{K}$ RLV subsets listed in  {Tab.~S2~\cite{SM}} (c1)-(c3) GL OPPs obtained from analytical minimization of the free-energy functionals. The $z^*$ coordinate is measured relative to the mean fluctuating CMI position $\langle \xi^*(t^*) \rangle$, where $z^* < 0$ corresponds to the melt and $z^* > 0$ corresponds to the crystal.}
\label{fig2}
\end{figure*}

\begin{figure*}[!htb]
\centering
\includegraphics[width=.99\textwidth]{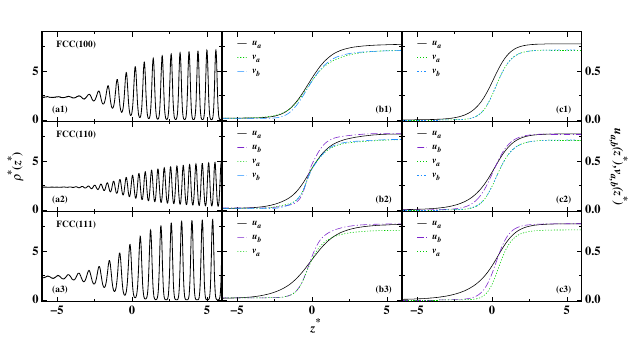}
\caption{(a1)-(a3) Fine-grained density profiles for the equilibrium FCC(100), FCC(110), and FCC(111) CMIs. (b1)-(b3) GL OPPs for the three CMIs computed from equilibrium MD simulations. (c1)-(c3) GL OPPs obtained from analytical minimization of the free-energy functionals.}
\label{fig3}
\end{figure*}

Panels (a1)-(a3) in Fig.~\ref{fig2} and Fig.~\ref{fig3} display the fine-scaled density variations across the three equilibrium BCC CMIs and three equilibrium FCC CMIs, respectively. The $z$-axis is centered at the fluctuating CMI position $\xi^*(t^*)$, with $z^* < 0$ corresponding to the melt region and $z^* > 0$ corresponding to the crystal region. The density oscillations are well-defined and orderly within the crystal phase, gradually diminishing as they transition smoothly into the melt phase along the $z$ axis.

The 10-90 interfacial widths ($\delta^{*}_{\rho}$) are estimated from the fine-grained density profiles as the distances over which the oscillation amplitudes transitions from 10\% to 90\% of their bulk crystal values.  The calculated $\delta^{*}_{\rho}$ values are approximately 3.08$\sigma$, 2.55$\sigma$, 2.33$\sigma$, 2.14$\sigma$, 2.10$\sigma$, and 2.09$\sigma$ for the BCC(100), (110), (111), FCC(100), (110), and (111) CMIs, respectively, as listed in Tab.~\ref{tab2}.

Panels (b1)-(b3) in Fig.\ref{fig2} and Fig.\ref{fig3} present complete sets of the GL order parameters computed from the MD simulation trajectories. The different curves in these panels correspond to various categories of the GL order parameters, which are based on the dot product of the CMI normal vector $\hat{n}$ and the  {RLVs} $\hat{K}$ (and $\hat{G}$). Within each subset, GL order parameters share identical density wave amplitudes across the CMI and are labeled as $u_a$, $u_b$, $\dots$ (and $v_a$, $\dots$), as shown in  {Tab.~S2 in the Supplemental Materials.~\cite{SM}} The single-mode GL model for BCC CMIs involves only $u_i$ ($i = a, b, \dots$), while the two-mode GL model for FCC CMIs involves both $u_i$ and $v_i$.

\begin{table*}[!htb]
\caption{10-90 widths of the fine-grained density profiles ($\delta^{*}_\rho$) and the GL OPPs ($\delta^{*}_{u,v}$) for BCC and FCC CMIs. The shifts of the GL OPP midpoints relative to that of the corresponding $u_a(z)$, $\Delta^{*}_{u,v} = z^{\mathrm{mid}*}_{u,v} - z^{\mathrm{mid}*}_{u_a}$. The superscripts “MD” and “AM” indicate the GL OPPs calculated from the MD simulation trajectory and those obtained from analytical minimization of the free-energy functionals, respectively.}
\begin{ruledtabular}
\begin{tabular}{rrrrrrrrrrrrrrrrrr}
BCC&
$\delta^{\mathrm{MD}{*}}_\rho$&
$\delta^{\mathrm{MD}{*}}_{u_a}$&
$\delta^{\mathrm{MD}{*}}_{u_b}$&
$\delta^{\mathrm{MD}{*}}_{u_c}$&
 &
$\Delta^{\mathrm{MD}{*}}_{u_a}$&
$\Delta^{\mathrm{MD}{*}}_{u_b}$&
$\Delta^{\mathrm{MD}{*}}_{u_c}$&
 &
$\delta^{\mathrm{AM}{*}}_{u_a}$&
$\delta^{\mathrm{AM}{*}}_{u_b}$&
$\delta^{\mathrm{AM}{*}}_{u_c}$&
 &
$\Delta^{\mathrm{AM}{*}}_{u_a}$&
$\Delta^{\mathrm{AM}{*}}_{u_b}$&
$\Delta^{\mathrm{AM}{*}}_{u_c}$&
\\
\hline
(100) & 3.08 & 3.57 & 3.91 & -\ \ \ &        & 0.00 & 0.03 & -\ \ \ & &  3.28 & 3.58 & -\ \ \ &             & 0.00 & -0.01 & -\ \ \  \\
(110) & 2.55 & 3.35 & 4.78 & 2.68 & & 0.00 & 0.01 & 0.02 & & 3.20 & 4.16 & 3.42 & & 0.00 & -0.03 & 0.00  \\
(111) & 2.33 & 3.82 & 2.58 & -\ \ \ &        & 0.00 & 0.03 & -\ \ \ & &  3.80 & 2.93 & -\ \ \ &             & 0.00 & 0.02 & -\ \ \  \\
\hline
FCC&
$\delta^{\mathrm{MD}{*}}_\rho$&
$\delta^{\mathrm{MD}{*}}_{u_a}$&
$\delta^{\mathrm{MD}{*}}_{u_b}$&
$\delta^{\mathrm{MD}{*}}_{v_a}$&
$\delta^{\mathrm{MD}{*}}_{v_b}$&
$\Delta^{\mathrm{MD}{*}}_{u_a}$&
$\Delta^{\mathrm{MD}{*}}_{u_b}$&
$\Delta^{\mathrm{MD}{*}}_{v_a}$&
$\Delta^{\mathrm{MD}{*}}_{v_b}$&
$\delta^{\mathrm{AM}{*}}_{u_a}$&
$\delta^{\mathrm{AM}{*}}_{u_b}$&
$\delta^{\mathrm{AM}{*}}_{v_a}$&
$\delta^{\mathrm{AM}{*}}_{v_b}$&
$\Delta^{\mathrm{AM}{*}}_{u_a}$&
$\Delta^{\mathrm{AM}{*}}_{u_b}$&
$\Delta^{\mathrm{AM}{*}}_{v_a}$&
$\Delta^{\mathrm{AM}{*}}_{v_b}$\\
\hline
(100) & 2.14 & 3.44 & -\ \ \ & 3.85 & 3.15 & 0.00 & -\ \ \ & -0.02 & 0.00 & 2.74 & -\ \ \ & 2.17 & 2.19 & 0.00 & -\ \ \ & 0.29 & 0.29  \\
(110) & 2.10 & 3.67 & 2.16 & 2.84 & 2.52 & 0.00 & -0.02 & -0.04 & -0.03 & 3.43 & 2.52 & 2.33 & 2.31 & 0.00 & 0.06 & 0.34 & 0.34 \\
(111) & 2.09 & 4.40 & 2.38 & 2.62 & -\ \ \ & 0.00 & 0.01 & -0.01 & -\ \ \  & 3.75 & 1.99 & 1.91 & -\ \ \  & 0.00 & 0.12 & 0.34 & -\ \ \  
\end{tabular}
\end{ruledtabular}
\label{tab2}
\end{table*}

The GL OPPs calculated in panels (b1)-(b3) show identical values of $u_i = u_\mathrm{s} = 0.722$ within the bulk crystal region for all three BCC CMI systems, and identical bulk crystal values of $u_i = u_\mathrm{s} = 0.779$ and $v_i = v_\mathrm{s} = 0.718$ for all three FCC CMI systems. A key difference between the GL order parameter values is that the density field around lattice points has smaller amplitude and is broader for BCC compared to FCC, reflecting the more closely packed nature of the latter. Following the approach of Wu \textit{et al.},~\cite{Wu16} we estimate the magnitude of the GL order parameters in the bulk crystals. Assuming the density field is represented as a sum of Gaussian peaks centered at lattice sites, the Fourier amplitudes of the density are simple functions of the corresponding wave numbers. The density field is given by: $\rho(\vec{r})=\rho_\mathrm{m}\left[1+\sum_{\vec{K}_{i}} \hat{\rho}_{\vec{K}_{i}}\mathrm{\bf{e}}^{i \vec{K}_{i} \cdot \vec{r}}+\sum_{\vec{G}_{i}} \hat{\rho}_{\vec{G}_{i}} \mathrm{\bf{e}}^{i \vec{G}_{i} \cdot \vec{r}}+ \cdots\right]$, where $\hat{\rho}_{\vec{K}_{i}}=\mathrm{\bf{e}}^{-\sigma^2_{\rho} {K}^2_{i}/2}$ and $\hat{\rho}_{\vec{G}_{i}}=\mathrm{\bf{e}}^{-\sigma^2_{\rho} {G}^2_{i}/2}$. The variance $\sigma^2_{\rho}$ of the Gaussian function is estimated by the mean-square displacements (MSD) of atoms in the crystals. Using MD data for the MSD of both BCC and FCC crystals at the corresponding $T_\mathrm{m}$, we estimate the magnitude of Fourier amplitudes $\hat{\rho}_{\vec{K}_{i}}$ and $\hat{\rho}_{\vec{G}_{i}}$ as approximations for $u_\mathrm{s}$ and $v_\mathrm{s}$. The estimates are $u^{\mathrm{est.}}_\mathrm{s}=\hat{\rho}_{\vec{K}_{i}}=0.716$ for BCC crystals, $u^{\mathrm{est.}}_\mathrm{s}=\hat{\rho}_{\vec{K}_{i}}=0.777$ and $v^{\mathrm{est.}}_\mathrm{s}=\hat{\rho}_{\vec{G}_{i}}=0.715$ for FCC crystals. These estimations are in good agreement with the values obtained from the GL OPPs in panels (b1)-(b3). Interestingly, similar estimations made for Ni metals are considerably smaller than the realistic values.~\cite{Wu16} On the bulk melt side (shown in panels (b1)-(b2) in Fig.~\ref{fig2} and Fig.~\ref{fig3}), the calculated amplitudes of density waves are sufficiently small, leading to a negligible contribution to the integrand function in the GL free-energy functionals  {[Eqs.(S12)-(S17) in Supplemental Materials.~\cite{SM}]}

The shapes of all calculated GL OPPs closely resemble hyperbolic tangent functions; however, the structural and directional dependence of their combinations varies significantly. To quantify this behavior, we measured the midpoint positions  $z^{\mathrm{mid}*}$ and the positions corresponding to 10\% and 90\% of the bulk crystal amplitudes $u_\mathrm{s}$ (and $v_\mathrm{s}$) on each calculated GL OPP. From the midpoint positions, we calculated the shift distance relative to the midpoint position of the corresponding $u_a(z)$ profile, defined as $\Delta^{*}_{u,v}=z^{\mathrm{mid}*}_{u,v}-z^{\mathrm{mid}*}_{u_a}$. As shown in Tab.~\ref{tab2}, most values of $|\Delta z^{*}_{u,v}|$ are around zero for the GL OPPs calculated from MD simulations. This indicates that the transitions of different density wave amplitude modes from the crystal to the melt phase occur in a spatially synchronous manner. The positions corresponding to 10\% and 90\% of the bulk crystal amplitudes were used to calculate the 10-90 interfacial widths $\delta^{*}_{u,v}$ for each GL OPP, which reveal the spatial decay rates of the density waves as they transition from the crystal to the melt. These widths may play a role in determining the anisotropy of the CMI $\gamma$. On the same CMI, the 10-90 interfacial width for the density profile $\delta^{*}_{\rho}$ is smaller than the 10-90 widths of the GL OPPs $\delta^{*}_{u,v}$, suggesting that the intrinsic density field containing purely particle-packing information across the CMI is less broader than any of the individual density wave amplitude profiles.

In general, BCC CMIs exhibit slightly larger interfacial widths than FCC CMIs. When comparing the three CMIs within each crystal structure (BCC or FCC), an interesting trend is observed for the (100) CMI. For both BCC and FCC, the 10-90 widths of the GL OPPs across different subsets are more tightly clustered compared to the (110) and (111) CMIs. This indicates that the (100) orientation has more uniform spatial decay rates of density wave amplitudes across all modes. Such behavior may underlie the observation that the (100) CMI has the highest value of $\gamma$ among the three orientations (as discussed in the next subsection).

For comparison, following the approach of Shih \textit{et al}.~\cite{Shih87} and Wu \textit{et al}.,~\cite{Wu06,Wu16} the GL OPPs for all CMIs are predicted using the analytical minimization (AM) method, as shown in panels (c1)-(c3) in Fig.~\ref{fig2} and Fig.~\ref{fig3}. Variational operations on the free energy functionals with respect to the corresponding order parameters lead to a system of coupled Euler-Lagrange (cE-L) equations. These equations are then solved numerically with boundary conditions: the order parameters vanish in the melt phase and retain fixed bulk solid values (we use  $u^{\mathrm{est.}}_\mathrm{s}$ and $v^{\mathrm{est.}}_\mathrm{s}$ as done by Wu \textit{et al}.~\cite{Wu16}). Additional details on the implementation of the analytical minimization of the free energy functionals can be found in the Supplemental Materials.~\cite{SM}

The GL OPPs predicted by the AM method and those measured from MD simulations show notable discrepancies. For the BCC CMIs, the GL OPPs predicted by the AM method agree well with the MD simulation results. The density wave amplitudes decay over similar spatial extents (see Tab.~\ref{tab2}) and follow a similar sequence. This good agreement is likely due to the alignment process [Eq.(\ref{eq:uvopro})], which helps eliminate capillary broadening effects in the GL OPP calculations, although capillary fluctuation effects are not considered in the AM method predictions.

In contrast, for the FCC CMIs, the GL OPPs predicted by the AM method differ significantly from those calculated by MD simulations. A distinct discrepancy is evident in Fig.~\ref{fig3}, where the GL OPPs predicted by the AM method show that $v_i(z)$ and $u_i(z)$ are not synchronous, with a shift distance $\Delta^*_v$ greater than 0.3$\sigma$. This contrasts with the MD simulation results, where the two modes are well synchronized. Such a spatial shift between the two sets of density wave amplitudes is unrealistic, as it contradicts the fact of a unified density field across the CMI. This discrepancy highlights the limitations of the AM method in providing accurate predictions for the two-mode GL model of FCC CMIs. We hypothesize that the issue arises from improper boundary conditions used during the analytical minimization, suggesting that additional constraints are needed to correctly enforce the relative spatial positions of the two sets of GL OPPs in FCC CMIs.

\subsection{Predicted $\gamma$ and the crystalline anistropy}

\begin{table*}[!htb]
\caption{The CMI $\gamma$ values and corresponding anisotropy parameters for both BCC and FCC CMIs are summarized. These are predicted using the GL models based on two sets of GL OPPs: those determined from MD simulations and those predicted via the AM method. Both sets of predictions are compared with the values measured in the simulation experiments conducted by Davidchack and Laird.~\cite{Davidchack05}}
\begin{ruledtabular}
\begin{tabular}{llllllll}
&$\gamma^*_{100}$ & $\gamma^*_{110}$ & $\gamma^*_{111}$ & $\gamma^*_{0}$  &  $\varepsilon_1$ & $\varepsilon_2$ & $\varepsilon_4$ \\
\hline
Sim. Exp.~\cite{Davidchack05}, BCC &0.642(8) &0.607(8) &0.620(7)&0.620(4) &0.06(3) &0.019(7) & 0.028(9) \\
GL, OPPs(MD), BCC &0.557 &0.539 &0.526&0.541 &0.082 & -0.006 & 0.016 \\
GL, OPPs(AM), BCC &0.537 &0.532 &0.519&0.530 &0.045 & -0.009 & 0.005\\
Sim. Exp.~\cite{Davidchack05}, FCC &0.873(8) &0.796(9) &0.764(11)&0.810(5) &0.20(2) &-0.003(8) & 0.046(7)\\
GL, OPPs(MD), FCC &0.919 &0.843 &0.827&0.861 &0.163 & 0.004 & 0.043\\
GL, OPPs(AM), FCC &0.726 &0.699 &0.681&0.702 &0.093 & -0.006 &0.019
\end{tabular}
\end{ruledtabular}
\label{tab3}
\end{table*}

The predicted values of $\gamma$ for three distinct orientations and two crystal structures, obtained from both the GL OPPs (derived from MD simulations) and the AM method, are summarized in Tab.~\ref{tab3}. These values are also compared with measurements by Davidchack and Laird using the cleaving method.~\cite{Davidchack05} To capture the weak capillary anisotropy of $\gamma$, a Kubic harmonics expansion~\cite{Altmann65,Fehlner76} is used. In this approach, the orientation-dependent of $\gamma$ is expressed as:
$\gamma^*_{\hat{n}}=\gamma^*_0+\varepsilon_1\gamma^*_0(\sum_i n_i^4-3/5)+\varepsilon_2\gamma^*_0(3\sum_i n_i^4+66n^2_xn^2_yn^2_z-17/7)$, 
where $\hat{n} = (n_x, n_y, n_z)$, $\gamma^*_0$ represents the mean value of $\gamma^*_{\hat{n}}$, and $\varepsilon_1$ and $\varepsilon_2$ characterize the capillary anisotropy. A positive $\varepsilon_1$ promotes dendrite growth along the $\langle 100 \rangle$ directions, while a negative $\varepsilon_2$ supports growth along the $\langle 110 \rangle$ directions. In addition to $\gamma^*_0$, $\varepsilon_1$, and $\varepsilon_2$, the anisotropy parameter $\varepsilon_4 = (\gamma^*_{100} - \gamma^*_{110}) / (\gamma^*_{100} + \gamma^*_{110})$, which is commonly used in dendrite growth theory, is also reported in Tab.~\ref{tab3}. This parameter provides an additional measure of the anisotropic behavior of $\gamma$ across orientations.

For BCC CMIs, the predicted values of $\gamma$ from the GL OPPs, derived using both MD simulations and the AM method, are in close agreement. The discrepancy between the two prediction methods for a given BCC CMI does not exceed 4\%. All predicted values cluster around $\gamma^*_0 = 0.54$ and 0.53, which are 12.8\% and 14.5\% lower, respectively, than the measurements obtained by Davidchack and Laird using simulation experiments.~\cite{Davidchack05} The predicted ordering of interfacial free energies follows the trend $\gamma^*_{111} < \gamma^*_{110} < \gamma^*_{100}$, which differs from the experimental results of $\gamma^*_{110} < \gamma^*_{111} < \gamma^*_{100}$. As a result, while the simple GL model provides a reasonable estimate for the magnitude of $\varepsilon_1$, it fails to accurately predict $\varepsilon_2$. In particular, the sign of $\varepsilon_2$ is opposite to that observed in the simulation experiment, indicating a limitation of the the simple GL model of BCC CMIs in capturing certain anisotropic features. Similar discrepancies have been reported in theoretical predictions by Wu and Karma,~\cite{Wu06,Wu07} who used both the single-mode GL (and PFC) model to estimate the BCC Fe CMI $\gamma$. Their results also showed an underestimation of approximately 20\% compared to MD simulation experiment data.~\cite{Sun04} 

For FCC CMIs, predictions based on GL OPPs determined from MD simulations show good agreement with the measurement data of Davidchack and Laird, with a difference in $\gamma^*_0$ of less than 6\%, indicating the physical plausibility of these predictions. In contrast, the GL model using GL OPPs determined via the AM method performs worse, underestimating the simulation experiment values by 14\%. At all CMI orientations, the predicted interfacial free energies for FCC CMIs are significantly higher than those for BCC CMIs. Both prediction methods correctly capture the ordering of interfacial free energies as $\gamma^*_{111} < \gamma^*_{110} < \gamma^*_{100}$, which aligns with the experimental results. The two-mode GL model, combined with MD simulation-based GL OPPs, provides reasonably accurate predictions for the anisotropy parameters $\varepsilon_1$, $\varepsilon_2$, and $\varepsilon_4$.
Notably, the predicted magnitudes of $\varepsilon_1$ and $\varepsilon_2$ are in perfect agreement with the values reported by Davidchack and Laird for the SS ($n=7$ and $n=8$) system,~\cite{Davidchack05} and they are closer to the measurements obtained for the SS ($n=6$) system compared to the predictions made using GL OPPs derived from the AM method. However, when using GL OPPs determined by the AM method, Wu \textit{et al.} applied two-mode and three-mode GL models to predict $\gamma$ for FCC Ni CMIs,~\cite{Wu16} but these predictions significantly underestimated the MD simulation experiment data~\cite{Hoyt03} by over 60\%. Additionally, the AM-based predictions failed to reproduce the ordering of $\gamma$ observed in the MD simulation experiment, highlighting the limitations of this approach for FCC CMIs.

The GL model predictions using GL OPPs derived from the AM method underestimate the simulation experiment measurements by approximately 14\% for both BCC and FCC CMI systems. However, the underlying reasons for these discrepancies differ between the two systems. 

For BCC CMIs, the discrepancy likely arises from the intrinsic limitations of the weakly nonlinear GL model, which employs a single set of density waves associated with the principal reciprocal lattice vectors (RLVs). This approach fails to capture the highly nonlinear nature of the crystal density field, which requires contributions from higher-order RLV modes for an accurate representation. Two key pieces of evidence support this explanation: (i) PFC simulations: Wu \textit{et al.}~\cite{Wu07} conducted PFC simulations that provided predictions of $\gamma$ that are in closer agreement with experimental measurements. the PFC simulation includes contributions from larger RLV modes, capturing the complex variations in the CMI density field that are neglected in the single-mode GL model. (ii)Similar underestimation with MD-based OPPs: The fact that the GL model predictions with MD-based GL OPPs exhibit a similar underestimation for BCC CMIs suggests that the discrepancy is rooted in the model's inherent limitations, rather than the method used to determine the OPPs. This indicates that the single-mode GL model's inability to account for contributions from higher-order RLV modes is a fundamental constraint that affects predictions regardless of how the OPPs are obtained.

For FCC CMIs, the source of the discrepancy appears to be different. When using MD-based GL OPPs, the predictions from the two-mode GL model (where $u_i(z)$ and $v_i(z)$ are synchronized) are significantly more accurate than those based on AM-derived GL OPPs, where $u_i(z)$ and $v_i(z)$ are desynchronized. This indicates that the primary cause of the discrepancy lies in the variational procedure used to derive the free energy functionals. The AM method may introduce inaccuracies by improperly handling the two sets of density waves, leading to a misrepresentation of the underlying physics. 

Addressing these issues requires two key improvements. (i) Incorporation of Larger RLV modes for BCC CMIs: expanding the GL model to include larger RLV modes could provide a more accurate depiction of the highly nonlinear density field in BCC CMIs. (ii) Optimization of variational methods for FCC CMIs: refining the variational approach used to derive the free energy functional in the two-mode GL model could eliminate the discrepancy for FCC CMIs. This may involve ensuring proper synchronization of the density wave components [$u_i(z)$ and $v_i(z)$] or adopting alternative analytical minimization methodologies. These enhancements have the potential to significantly reduce the observed discrepancies and improve the predictive accuracy of the GL model for both BCC and FCC CMIs.

\begin{figure}[!htb]
\centering
\includegraphics[width=0.45\textwidth]{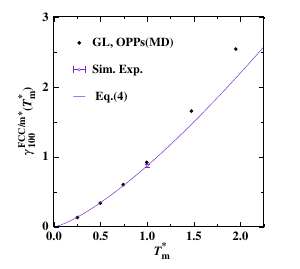}
\caption{The GL model predictions of $\gamma^\mathrm{FCC/m*}_{100}$ with the MD-based GL OPPs (filled circles), compared with computer simulation experimental value ($\gamma^\mathrm{FCC/m*}_{1,100}=0.873$ at $T_\mathrm{m}=1.0$, open circle) reported by Davidchack and Laird.~\cite{Davidchack05} and  the SS scaling law of $\gamma$  (solid line) as described with Eq.(\ref{eq.scaling.gamma}), i.e., $\gamma^\mathrm{FCC/m*}_{100}=0.873T_\mathrm{m}^{*4/3}$ for SS ($n=6$) system.}
\label{fig4}
\end{figure} 

\begin{table*}[!htb]
\caption{The contribution components of the $\gamma^*$, the component $\gamma^*_\mathrm{nonSG}$ arises from the non-square gradient term and the other component(s) arise from the square gradient terms ($\gamma^*_{u\mathrm{SG}}$ and/or $\gamma^*_{v\mathrm{SG}}$) of the GL OPPs due to the principal and/or the second set of RLVs.  {See the detailed expression for $\gamma^*_\mathrm{nonSG}$ and others in Eq.(S18) in the Supplementary Materials.}}
\begin{ruledtabular}
\begin{tabular}{lrclrclrclrcl}
&
\multicolumn{6}{c}{------  \          GL prediction from OPPs(MD)       \ ------}&
\multicolumn{6}{c}{------ \       GL prediction from OPPs(AM)       \ ------}\\
&BCC (100)&(110)&(111)   \ \ &FCC (100)&(110)&(111)   \ \ &BCC (100)&(110)&(111)  \  \ &FCC (100)&(110)&(111)   \  \ \\
\hline
$\gamma^*$  & 0.557 & 0.539 & 0.526 & 0.919 & 0.843 & 0.827 & 0.537 & 0.532 &  0.519 & 0.726 & 0.699 & 0.681  \\
$\gamma^*_{\mathrm{nonGS}}$  & 0.346 & 0.328 & 0.304 & 0.645 & 0.578 & 0.540 & 0.301 & 0.297 &  0.296 & 0.364 & 0.401 & 0.340  \\
$\gamma^*_{u\mathrm{GS}}$   & 0.211 & 0.211 & 0.222 & 0.252 & 0.237 & 0.256 & 0.236 & 0.235 &  0.223 & 0.328 & 0.267 & 0.301  \\
$\gamma^*_{v\mathrm{GS}}$    & - & - & - & 0.022 & 0.029 & 0.032  & - & - & - & 0.034 & 0.032 & 0.039 \\
\end{tabular}
\end{ruledtabular}
\label{tab4}
\end{table*}

As demonstrated above, the predictions of $\gamma$ and its anisotropy for FCC CMIs, using MD simulation-based GL OPPs and the two-mode GL model, closely match the simulation experiment results. To the best of our knowledge, this level of agreement surpasses all prior studies that have employed either GL or PFC models. A follow-up question naturally arises: What causes the observed $\sim 6\%$ overestimation? To investigate this, we leverage an important feature of the SS model system (introduced in subsection~\ref{subsec-ss}) as a diagnostic tool. Specifically, we compare the GL model predictions along the $p$-$T$ coexistence boundary with the power-law scaling of the SS ($n=6$) system for the CMI $\gamma$, as described by Eq.(\ref{eq.scaling.gamma}). This approach provides insight into the origin of the overestimation, allowing for a deeper understanding of the underlying mechanisms that contribute to the observed deviation.

Using FCC(100) as a representative CMI system, we conduct equilibrium MD simulations at six different coexistence conditions along the $p$-$T$ coexistence boundary. As illustrated in Fig.~\ref{fig4}, the GL model predictions of $\gamma^\mathrm{FCC/m*}_{100}$, obtained using the calculated GL OPPs and the multiplicative coefficients in  {Eq.(S15),~\cite{SM}} show excellent agreement with the trend predicted by the scaling law [Eq.(\ref{eq.scaling.gamma})]. This agreement persists over a wide temperature range up to approximately 0.75. However, a progressive overestimation of $\gamma^\mathrm{FCC/m*}_{100}$ is observed as the temperature increases beyond this point, suggesting a temperature-dependent deviation that warrants further investigation.

An inspection of the key contributing parameters in the GL model expression for $\gamma^\mathrm{FCC/m*}_{100}$  {[Eq.(S15)]~\cite{SM}} provides a clue to the possible origin of the overestimation. Fig.S1 in the Supplemental Materials shows that the densities of the uniform bulk melt phases, $\rho^*_\mathrm{m}$, at crystal-melt coexistence conditions (temperature and pressure) follow the expected power-law scaling. The variation in $\gamma^\mathrm{FCC/m*}_{100}/\rho_\mathrm{m}$ with respect to the dimensionless pressure ($p^*$) or melting temperature ($T_{\mathrm{m}}^{*}$) is governed by the product of two key factors: (1) Multiplicative coefficients: these coefficients are linked to liquid structure factors or direct correlation functions, as summarized in  {Tab.~S1.~\cite{SM}} (2) GL OPP-dependent terms: these include power and gradient terms that explicitly depend on the GL OPPs. Interestingly, it is observed that both (i) the structure factors (or direct correlation functions) in terms of dimensionless reduced wave numbers $k^*a^*$ for the bulk melt phases, and (ii) the calculated GL OPPs for the six FCC(100) CMIs along the dimensionless reduced $z^*/a^*$ coordinate, exhibit identical shapes that are independent of pressure and temperature, as shown in Fig.S2 and Fig.S3 of the Supplemental Materials.

The invariant GL OPPs, calculated from CMI atomistic trajectories, serve as exact and valid input data for the GL model. Therefore, we speculate that the observed deviations in predicting $\gamma^\mathrm{FCC/m*}_{100}$ at higher $p^*$ or $T_{\mathrm{m}}^{*}$ are primarily associated with assumptions made in the formalism of the CMI free energy functionals. These assumptions include: (i) Constant RLVs across the CMI: the reciprocal lattice vectors (RLVs) are assumed to remain constant throughout the CMI, which may oversimplify the actual structural changes occurring at the interface. (ii)Bulk liquid structure factor for interfacial liquid: the free energy functional relies on multiplicative coefficients related to the bulk liquid structure factors, implicitly assuming that the interfacial liquid has the same structural properties as the bulk melt phase. Support for this speculation comes from computational evidence indicating that the local structure factor of interfacial liquids differs from that of the corresponding bulk liquid phase.~\cite{Miyahara97} Such differences highlight the limitations of the current GL model's treatment of interfacial liquid structure. 

While the current GL model benefits from computational simplicity, further improvements could be achieved by incorporating more detailed structural variations into the model. An updated version of the GL model could account for local deviations in structure factors and RLVs within the CMI region. Such refinements would likely enhance the quantitative accuracy of predictions for $\gamma^*$, especially at higher $p^*$ and $T_{\mathrm{m}}^{*}$.

\begin{figure}[!htb]
\centering
\includegraphics[width=0.45\textwidth]{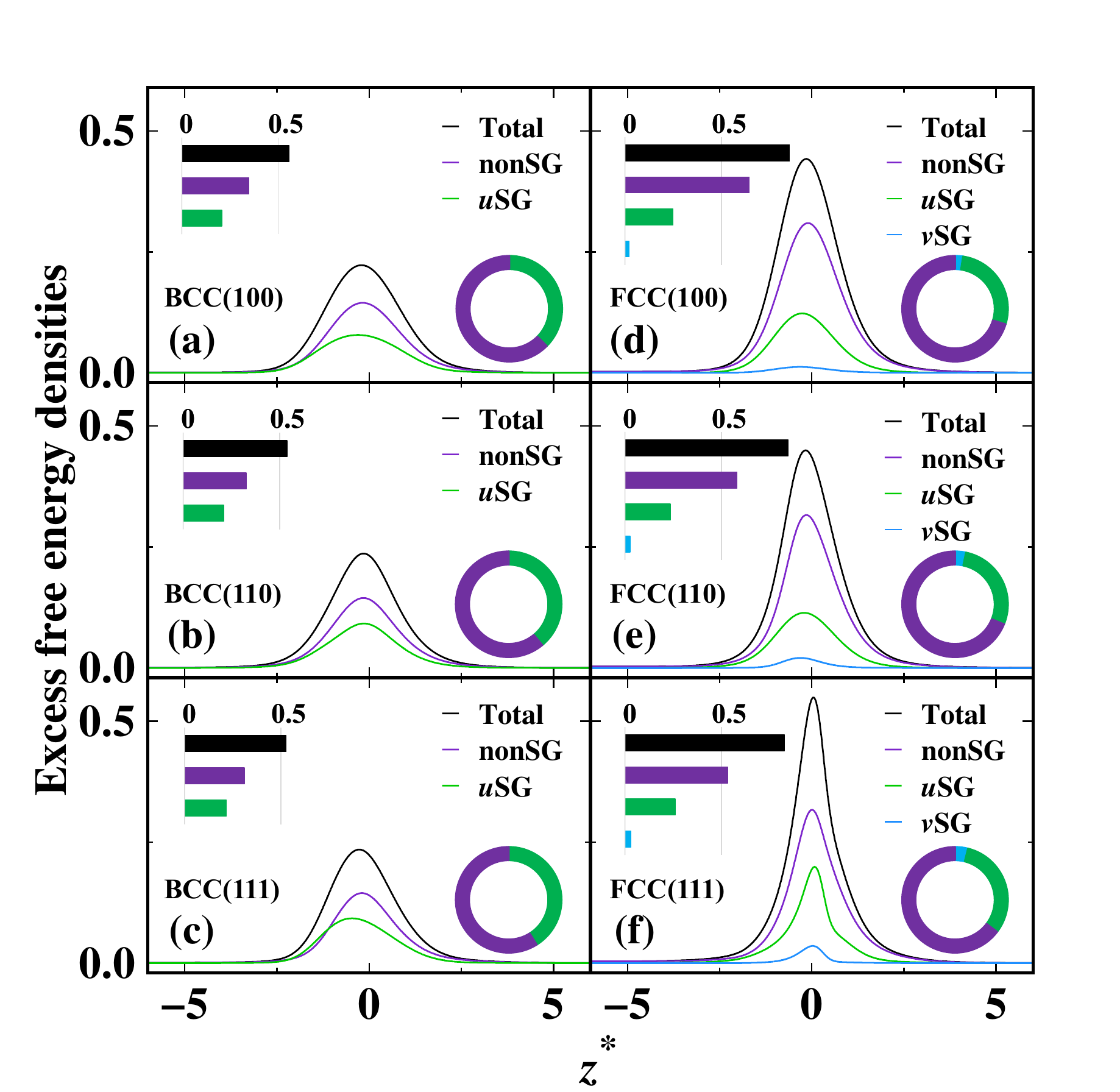}
\caption{Excess free energy densities (black solid lines) for the BCC (left side) and FCC (right side) CMIs as a function of distance $z^*$. The individual contribution components are also visualized using colored lines, including non-square gradient (nonSG) terms and square gradient ($u$SG, or $v$SG) terms.}
\label{fig5}
\end{figure}

We present key insights into the crystal structure dependency and crystalline anisotropy of $\gamma$ using the GL model, which demonstrates strong predictive capability. The interfacial free energy $\gamma^*$ is decomposed into contributions from non-square gradient ($\gamma^*_{\mathrm{nonSG}}$) and square gradient terms ($\gamma^*_{u\mathrm{SG}}$ and $\gamma^*_{v\mathrm{SG}}$), associated with the principal and second set of reciprocal lattice vectors (RLVs), respectively,  {see the detailed expression for $\gamma^*_\mathrm{nonSG}$, $\gamma^*_{u\mathrm{SG}}$ and $\gamma^*_{v\mathrm{SG}}$ in Eq.(S18) in the Supplementary Materials.} Data derived from MD-based OPPs in Tab.~\ref{tab4} and the integrands of the CMI free energy functionals in Fig.~\ref{fig5} illustrate this decomposition. The overall interfacial free energy $\gamma^{\mathrm{BCC/m*}}$ for BCC crystal-melt interfaces (CMIs) is lower than $\gamma^{\mathrm{FCC/m}*}$ for FCC CMIs, primarily due to the larger contribution of non-gradient terms, as evidenced in Fig.~\ref{fig5} and Tab.~\ref{tab4}. This highlights a potential area for future refinement of the GL model for BCC CMIs, particularly by incorporating additional RLV modes to improve accuracy. Our analysis reveals that non-gradient terms are the dominant contributors to $\gamma^*$ and the primary drivers of its anisotropy. By contrast, square gradient terms have minimal impact on anisotropy. For both BCC and FCC CMIs, the values of $\gamma^*_{u\mathrm{SG}}$ across the three crystallographic orientations are nearly identical, while $\gamma^*_{v\mathrm{SG}}$ for FCC CMIs is approximately one-tenth the magnitude of $\gamma^*_{u\mathrm{SG}}$. Additionally, FCC CMIs exhibit larger anisotropy compared to BCC CMIs. 

These observations suggest that anisotropy can be modulated by adjusting the relative contributions of non-gradient and square gradient terms. One example, illustrated in Fig.~S4 of the Supplementary Materials, shows that reducing pressure increases the relative weight of square gradient terms, which may offer a strategy to suppress capillary anisotropy. This finding underscores the potential for tailoring anisotropy through the careful tuning of GL model parameters and structural inputs.

\section{Summary}

We propose a new method for quantitatively predicting the crysta-melt interfacial free energy $\gamma$ within the GL model framework by incorporating atomic-scale simulation data to generate more accurate density wave profiles. The study focuses on the soft-sphere (SS) system modeled by the inverse power potential with $n=6$. By combining MD-based density wave profiles with existing GL models for both BCC and FCC CMIs, we achieve accurate predictions for $\gamma$ and its anisotropy. Notably, for FCC CMIs, the level of agreement obtained in this study surpasses all prior predictions using either GL or PFC models. Our findings establish that the GL model is both valid and reliable, offering a computationally efficient and cost-effective approach to estimating CMI $\gamma$ with reasonable accuracy.

This study also illustrates how the GL model can be employed to interpret the crystal structure dependency and anisotropy of $\gamma$ by leveraging the contributions of different components in the analytical expression for $\gamma$. By comparing results from the conventional approach (where free energy functionals are analytically minimized) with predictions based on the power-law scaling of SS CMI $\gamma$ along the $p$-$T$ coexistence boundary, we obtain several key insights: {(i)} For the $\gamma$ of BCC CMIs, the current single-mode GL model requires additional Fourier components of the density and higher-order expansion terms in the free energy functional to achieve higher accuracy. (ii) For FCC systems, the larger underestimation of $\gamma$ when using GL order parameter profiles (OPPs) from the AM method is attributed to improper variational operations on the free energy functional. An optimized variational procedure is necessary for better predictions. (iii) To achieve more quantitative predictions, future models should incorporate detailed structural information for interfacial liquids rather than relying on bulk melt phase data for structure factors. This adjustment could significantly enhance the precision of the GL model for CMIs.

To conclude, this study provides new perspectives on advancing and refining the GL model for predicting CMI $\gamma$. A future, more accurate version of the GL model could facilitate a deeper analytical understanding of $\gamma$ and the mechanisms driving various physical phenomena. Potential applications include the elucidation of the effects of material (or interatomic interactions), crystal structural (e.g., non-cubic crystal structure under strain, or HCP, etc.), composition dependencies and so on.


\begin{acknowledgments}
YY acknowledges the Chinese National Science Foundation (Grant No. 11874147), the Natural Science Foundation of Shanghai (Grant No. 23ZR1420200).
\end{acknowledgments}

\bibliography{GM}
\end{document}


\title{Supplemental Materials ``A quantitative theory and atomistic simulation study on the soft-sphere crystal-melt interfacial properties: II. Interfacial free energies''}

\author{Ya-Shen Wang}
\affiliation{State Key Laboratory of Precision Spectroscopy, School of Physics and Electronic Science, East China Normal University, Shanghai 200241, China}
\author{Zun Liang}
\affiliation{State Key Laboratory of Precision Spectroscopy, School of Physics and Electronic Science, East China Normal University, Shanghai 200241, China}
\author{Brian B. Laird}
\thanks{blaird@ku.edu}
\affiliation{Department of Chemistry, University of Kansas, Lawrence, KS 66045, USA}
\affiliation{Freiburg Institute for Advanced Studies (FRIAS), Albert-LudwiSG-Universit\"{a}t Freiburg, Albertstra{\ss}e 19, 79104 Freiburg im Breisgau, Germany}
\author{Yang Yang}
\thanks{yyang@phy.ecnu.edu.cn}
\affiliation{State Key Laboratory of Precision Spectroscopy, School of Physics and Electronic Science, East China Normal University, Shanghai 200241, China}

\maketitle



\noindent {\bf TABLE OF CONTENTS}

\vspace{0.5cm}
\noindent {\bf I. Theory Formlism}

\vspace{0.5cm}
{\bf A. Density functionals}

\vspace{0.5cm}
{\bf B. Free energy functionals}

\vspace{0.5cm}
{\bf C. GL model formlism for specific CMI orientations}

\vspace{0.5cm}
\noindent {\bf II. Analytically minimization of the free energy functionals}

\vspace{0.5cm}
\noindent {\bf III. Supplemental table}

\vspace{0.5cm}
\noindent {\bf IV. Supplemental figures}

\newpage

\vspace{1.5cm}
\noindent {\bf I. Theory Formlism}
\label{sec:ap3}
\vspace{.3cm}

\noindent {\bf A. Density functionals}
\label{sec:ap3}
\vspace{.3cm}

For the BCC CMIs, the density field $\rho^\mathrm{BCC/m}(\vec{r})$ is expanded and truncated up to the contribution of the principal reciprocal-lattice vectors (RLVs), $\vec{K}_{i}$, negelecting the contribution of larger RVL modes,
\begin{equation}
\rho^\mathrm{BCC/m}(\vec{r})=\rho_\mathrm{m}\left[1+\sum_{\vec{K}_{i}} u_{i}(\vec{r}) \mathrm{\bf{e}}^{i \vec{K}_{i} \cdot \vec{r}} \right],
\label{eq:dens-BCC}
\end{equation}
where the order parameters $u_i$ are the density waves amplitudes corresponding to the twelve $\left\langle110\right\rangle$ principal RLVs of equal length, in a BCC crystal. $\rho_\mathrm{m}$ is the density of the uniform bulk melt phase at the crystal-melt coexistence temperature (or melting temperature $T_\mathrm{m}$) and pressure.

For the FCC CMIs, the density field $\rho^\mathrm{FCC/m}(\vec{r})$ is expanded and truncated up to the second set of RLVs, $\vec{G}_{i}$,
\begin{equation}
\rho^\mathrm{FCC/m}(\vec{r})=\rho_\mathrm{m}\left[1+\sum_{\vec{K}_{i}} u_{i}(\vec{r}) \mathrm{\bf{e}}^{i \vec{K}_{i} \cdot \vec{r}}+\sum_{\vec{G}_{i}} v_{i}(\vec{r}) \mathrm{\bf{e}}^{i \vec{G}_{i} \cdot \vec{r}}\right],
\label{eq:dens-FCC}
\end{equation}
where the order parameters $u_i$ are the density waves amplitudes corresponding to the eight $\left\langle111\right\rangle$ principal RLVs of equal length, in a FCC crystal. $v_i$ are the density waves amplitudes corresponding to the six $\left\langle200\right\rangle$ second set of RLVs of equal length, in a FCC crystal.

\label{sec:ap3}
\vspace{.3cm}
\noindent {\bf B. Free energy functionals}
\label{sec:ap3}
\vspace{.3cm}

The free-energy functionals in GL models of the BCC and FCC CMIs are firstly derived from the classical density functional theory (cDFT) of freezing,~\cite{Haymet81,Oxtoby82} considering inhomogeneous fluids with small density fluctuations. The excess free-energy $\Delta F$ for the equilibrium CMIs relative to the free energy of the uniform melt phase $F_\mathrm{m}$, $\Delta F \equiv F-F_\mathrm{m}$, is expressed as, 
\begin{small}
\begin{equation}
\Delta F =\frac{1}{2\beta_\mathrm{m}}\iint\mathrm{d}\vec{r}\mathrm{d}\vec{r \ }'\delta\rho(\vec{r})\left[\frac{\delta(\vec{r}-\vec{r \ }')}{\rho_\mathrm{m}}-C(|\vec{r}-\vec{r \ }'|)\right]\delta\rho(\vec{r \ }'),
\label{eq:DeltaF-cdft}
\end{equation}
\end{small}
in which, $\beta_\mathrm{m}=1/{k_\mathrm{B}T_\mathrm{m}}$. $\delta\rho(\vec{r})\equiv \rho(\vec{r})-\rho_\mathrm{m}$, $C(|\vec{r}-\vec{r \ }'|)$ is the two-particle direct correlation function of the melt phase. 

We parametrize the direction normal to the planar CMIs is along $z$ direction. By assuming that the amplitudes of density waves vary slowly across the CMIs, the density fluctuations, for both the BCC CMIs [$\delta\rho^\mathrm{BCC/m}(\vec{r \ }')$] and FCC CMIs [$\delta\rho^\mathrm{FCC/m}(\vec{r \ }')$], can be expanded in a Taylor series about $z$,~\cite{Wu06,Wu16}

\begin{equation}
\begin{aligned}
\delta\rho^\mathrm{BCC/m}(\vec{r \ }') \approx &\rho_\mathrm{m} \Bigg\{ \sum_{\vec{K}_{i}}
\bigg[ u_{i}(z) + \frac{\mathrm{d}u_{i}(z)}{\mathrm{d}z} (z'-z) + \frac{1}{2}\frac{\mathrm{d}^2u_{i}(z)}{\mathrm{d}z^2} (z'-z)^2 \bigg]\mathrm{\bf{e}}^{i \vec{K}_{i} \cdot \vec{r}'}\Bigg\} ,
\label{eq:dens-expan-BCC}
\end{aligned}
\end{equation}

\begin{equation}
\begin{aligned}
\delta\rho^\mathrm{FCC/m}(\vec{r \ }') \approx &\rho_\mathrm{m} \Bigg\{ \sum_{\vec{K}_{i}}
\bigg[ u_{i}(z) + \frac{\mathrm{d}u_{i}(z)}{\mathrm{d}z} (z'-z) + \frac{1}{2}\frac{\mathrm{d}^2u_{i}(z)}{\mathrm{d}z^2} (z'-z)^2 \bigg]\mathrm{\bf{e}}^{i \vec{K}_{i} \cdot \vec{r}'}\Bigg\}  \\
+&\rho_\mathrm{m} \Bigg\{ \sum_{\vec{G}_{i}}
\bigg[ v_{i}(z) + \frac{\mathrm{d}v_{i}(z)}{\mathrm{d}z} (z'-z) + \frac{1}{2}\frac{\mathrm{d}^2v_{i}(z)}{\mathrm{d}z^2} (z'-z)^2 \bigg]\mathrm{\bf{e}}^{i \vec{G}_{i} \cdot \vec{r}'}\Bigg\} 
\label{eq:dens-expan-FCC}
\end{aligned}
\end{equation}

The excess free-energies for the equilibrium BCC CMIs and FCC CMIs are calculated by substituting the density fluctuations [Eq.(\ref{eq:dens-expan-BCC}) and Eq.(\ref{eq:dens-expan-FCC})] in Eq.(\ref{eq:DeltaF-cdft}) and carry out the spatial integration over $r'$, leading to, 
\begin{small}
\begin{equation}
\begin{aligned}
\Delta F^\mathrm{BCC/m} \approx & \frac{\rho_\mathrm{m}}{2\beta_\mathrm{m}}\int\mathrm{d}\vec{r} \Bigg(\sum_{\vec{K}_{i},\vec{K}_{j}}
\frac{1}{S(|\vec{K}_{i}|)}u_iu_j\delta_{0,\vec{K}_{i}+\vec{K}_{j}}
-\sum_{\vec{K}_{i}}\frac{1}{2}\bigg\{C^{\prime\prime}(|\vec{K}_{i}|)(\hat{K}_{i} \cdot \hat{z})^2 
+\frac{C^{\prime}(|\vec{K}_{i}|)}{|\vec{K}_{i}|}[1-(\hat{K}_{i} \cdot \hat{z})^2 ] \bigg\} \left| \frac{\mathrm{d}u_i}{\mathrm{d}z}\right|^2\Bigg),
\label{eq:DeltaF-cdft-BCC}
\end{aligned}
\end{equation}
\end{small}
\begin{small}
\begin{equation}
\begin{aligned}
\Delta F^\mathrm{FCC/m} \approx & \frac{\rho_\mathrm{m}}{2\beta_\mathrm{m}}\int\mathrm{d}\vec{r} 
\Bigg(\sum_{\vec{K}_{i},\vec{K}_{j}}
\frac{1}{S(|\vec{K}_{i}|)}u_iu_j\delta_{0,\vec{K}_{i}+\vec{K}_{j}} 
+\sum_{\vec{G}_{i},\vec{G}_{j}}
\frac{1}{S(|\vec{G}_{i}|)}v_iv_j\delta_{0,\vec{G}_{i}+\vec{G}_{j}} \\
-&\sum_{\vec{K}_{i}}\frac{1}{2}\bigg\{C^{\prime\prime}(|\vec{K}_{i}|)(\hat{K}_{i} \cdot \hat{z})^2 +\frac{C^{\prime}(|\vec{K}_{i}|)}{|\vec{K}_{i}|}[1-(\hat{K}_{i} \cdot \hat{z})^2 ] \bigg\} \left| \frac{\mathrm{d}u_i}{\mathrm{d}z}\right|^2\\
-&\sum_{\vec{G}_{i}}\frac{1}{2}\bigg\{C^{\prime\prime}(|\vec{G}_{i}|)(\hat{G}_{i} \cdot \hat{z})^2 +\frac{C^{\prime}(|\vec{G}_{i}|)}{|\vec{G}_{i}|}[1-(\hat{G}_{i} \cdot \hat{z})^2 ] \bigg\} \left| \frac{\mathrm{d}v_i}{\mathrm{d}z}\right|^2\Bigg),
\label{eq:DeltaF-cdft-FCC}
\end{aligned}
\end{equation}
\end{small}
in which, $C^{\prime}(k)=\mathrm{d}C(k)/\mathrm{d}k$ and $C^{\prime\prime}(k)=\mathrm{d}^2C(k)/\mathrm{d}k^2$, the hat symbols on top of $\hat{K}_{i}$, $\hat{G}_{i}$ and $\hat{z}$ stand for the unit vectors. $C(k)$ is the Fourier transform of $C(|\vec{r}|)$,
\begin{equation}
C(k)=\rho_\mathrm{m}\int \mathrm{d}\vec{r}C(|\vec{r}|)\mathrm{\bf{e}}^{i \vec{k} \cdot \vec{r}},
\label{eq:ck}
\end{equation}
and it is related to the structure factor of the melt phase, 
\begin{equation}
C(k)=\frac{S(k)-1}{S(k)}.
\label{eq:ck-sk}
\end{equation}

In the following theoretical formlism, we follow the assumptions that $C^{\prime}(|\vec{K}_{i}|)=0$ for both the BCC and FCC CMI systems,~\cite{Wu06,Wu15,Wu16} as the principal RLVs $|\vec{K}_{i}|$ of BCC and FCC crystals lie closest to the maximum in the liquid structure factor, whereas, for the second set of RLVs $|\vec{G}_{i}|$, the FCC CMI systems, the first derivative of $C(k)$ does not vanish. At the same time, the RLVs are assumed as constant across the CMIs, the ratio of the lattice spacing (e.g., under expansion) to the spatially diffuse width (or the correlation length of the melt phase) of the atomically rough CMI is treated as a small parameter.~\cite{Wu15}

To complete the formlism of the GL model of the free-energy functionals of the equilibrium BCC and FCC CMIs, after obtaining the standard cDFT expression for the free-energy functionals, i.e., Eq.(\ref{eq:DeltaF-cdft-BCC}) and Eq.(\ref{eq:DeltaF-cdft-FCC}), one expand the excess free energies as power series of the density wave amplitudes (or the GL order parameters),

\begin{equation}
\begin{aligned}
\Delta F^\mathrm{BCC/m} \approx &\frac{\rho_\mathrm{m}}{2\beta_\mathrm{m}}\int\mathrm{d}\vec{r}  
\Bigg(a^\mathrm{BCC/m}_{uu}\sum_{\vec{K}_{i},\vec{K}_{j}}C^\mathrm{BCC/m}_{uu}u_iu_j\delta_{0,\vec{K}_{i}+\vec{K}_{j}} 
-a^\mathrm{BCC/m}_{uuu}\sum_{\vec{K}_{i},\vec{K}_{j},\vec{K}_{k}}C^\mathrm{BCC/m}_{uuu}u_iu_ju_k\delta_{0,\vec{K}_{i}+\vec{K}_{j}+\vec{K}_{k}}\\
+&a^\mathrm{BCC/m}_{uuuu}\sum_{\vec{K}_{i},\vec{K}_{j},\vec{K}_{k},\vec{K}_{l}}C^\mathrm{BCC/m}_{uuuu}u_iu_ju_ku_l\delta_{0,\vec{K}_{i}+\vec{K}_{j}+\vec{K}_{k}+\vec{K}_{l}}
+a^\mathrm{BCC/m}_{u\mathrm{SG}}\sum_{\vec{K}_{i}}C^\mathrm{BCC/m}_{u\mathrm{SG}}\left| \frac{\mathrm{d}u_i}{\mathrm{d}z}\right|^2\Bigg),
\label{eq:DeltaF-GL-BCC}
\end{aligned}
\end{equation}

\begin{equation}
\begin{aligned}
\Delta F^\mathrm{FCC/m} \approx &\frac{\rho_\mathrm{m}}{2\beta_\mathrm{m}}\int\mathrm{d}\vec{r}  
\Bigg(a^\mathrm{FCC/m}_{uu}\sum_{\vec{K}_{i},\vec{K}_{j}}C^\mathrm{FCC/m}_{uu}u_iu_j\delta_{0,\vec{K}_{i}+\vec{K}_{j}} 
+a^\mathrm{FCC/m}_{vv}\sum_{\vec{G}_{i},\vec{G}_{j}}C^\mathrm{FCC/m}_{vv}v_iv_j\delta_{0,\vec{G}_{i}+\vec{G}_{j}} \\
-&a^\mathrm{FCC/m}_{uuv}\sum_{\vec{K}_{i},\vec{K}_{j},\vec{G}_{k}}C^\mathrm{FCC/m}_{uuv}u_iu_jv_k\delta_{0,\vec{K}_{i}+\vec{K}_{j}+\vec{G}_{k}}
+a^\mathrm{FCC/m}_{uuuu}\sum_{\vec{K}_{i},\vec{K}_{j},\vec{K}_{k},\vec{K}_{l}}C^\mathrm{FCC/m}_{uuuu}u_iu_ju_ku_l\delta_{0,\vec{K}_{i}+\vec{K}_{j}+\vec{K}_{k}+\vec{K}_{l}} \\
+&a^\mathrm{FCC/m}_{uuvv}\sum_{\vec{K}_{i},\vec{K}_{j},\vec{G}_{k},\vec{G}_{l}}C^\mathrm{FCC/m}_{uuvv}u_iu_jv_kv_l\delta_{0,\vec{K}_{i}+\vec{K}_{j}+\vec{G}_{k}+\vec{G}_{l}} 
+a^\mathrm{FCC/m}_{u\mathrm{SG}}\sum_{\vec{K}_{i}}C^\mathrm{FCC/m}_{u\mathrm{SG}}\left| \frac{\mathrm{d}u_i}{\mathrm{d}z}\right|^2\\
+&\sum_{\vec{G}_{i}}(a^\mathrm{FCC/m}_{v\mathrm{SG}-L} C^\mathrm{FCC/m}_{v\mathrm{SG}-L}+a^\mathrm{FCC/m}_{v\mathrm{SG}-T} C^\mathrm{FCC/m}_{v\mathrm{SG}-T})\left| \frac{\mathrm{d}v_i}{\mathrm{d}z}\right|^2 \Bigg),
\label{eq:DeltaF-GL-FCC}
\end{aligned}
\end{equation}

\begin{table*}[!htb]
\caption{List of the coefficients in the free-energy functional expressions for the BCC CMIs Eq.(\ref{eq:DeltaF-GL-BCC}) and FCC CMIs and Eq.(\ref{eq:DeltaF-GL-FCC}). The subscripts `s' in `$u_\mathrm{s}$' and `$v_\mathrm{s}$' stands for the magnitude of the density wave amplitudes in bulk solid (crystal) phase.}
\begin{ruledtabular}
\begin{tabular}{llll}
\multicolumn{2}{c}{BCC CMIs}&\multicolumn{2}{c}{FCC CMIs}\\
\hline
$a^\mathrm{BCC/m}_{uu}=\frac{12}{S(|\vec{K}_{110}|)}$ & $C^\mathrm{BCC/m}_{uu}=\frac{1}{12}$ & $a^\mathrm{FCC/m}_{uu}=\frac{8}{S(|\vec{K}_{111}|)}$ & $C^\mathrm{FCC/m}_{uu}=\frac{1}{8}$\\
 & & & \\
- & - & $a^\mathrm{FCC/m}_{vv}=\frac{6}{S(|\vec{G}_{200}|)}$    &$C^\mathrm{FCC/m}_{vv}=\frac{1}{6}$\\
 & & & \\
$a^\mathrm{BCC/m}_{uuu}=\frac{24}{[S(|\vec{K}_{110}|)u_\mathrm{s}^\mathrm{BCC/m}]}$ &  $C^\mathrm{BCC/m}_{uuu}=\frac{1}{8}$ &$a^\mathrm{FCC/m}_{uuv}=\frac{16}{S(|\vec{K}_{111}|)v_\mathrm{s}^\mathrm{FCC/m}}+\frac{12v_\mathrm{s}^\mathrm{FCC/m}}{S(|\vec{G}_{200}|)(u_\mathrm{s}^\mathrm{FCC/m})^2}$   &$C^\mathrm{FCC/m}_{uuv}=\frac{1}{12}$\\
 & & & \\
$a^\mathrm{BCC/m}_{uuuu}=\frac{12}{[S(|\vec{K}_{110}|)(u_\mathrm{s}^\mathrm{BCC/m})^2]}$ & $C^\mathrm{BCC/m}_{uuuu}=\frac{1}{27}$ &$a^\mathrm{FCC/m}_{uuuu}=\frac{6(v_\mathrm{s}^\mathrm{FCC/m})^2}{[S(|\vec{G}_{200}|)(u_\mathrm{s}^\mathrm{FCC/m})^4]}$  &$C^\mathrm{FCC/m}_{uuuu}=\frac{1}{12}$\\
 & & & \\
- & - &$a^\mathrm{FCC/m}_{uuvv}=\frac{8}{[S(|\vec{K}_{111}|)(v_\mathrm{s}^\mathrm{FCC/m})^2]}$  &$C^\mathrm{FCC/m}_{uuvv}=\frac{1}{24}$\\
 & & & \\
$a^\mathrm{BCC/m}_{u\mathrm{SG}}=-2C^{\prime\prime}(|\vec{K}_{110}|)$ & $C^\mathrm{BCC/m}_{u\mathrm{SG}}=\frac{(\hat{K}_i\cdot\hat{n})^2}{4}$ & $a^\mathrm{FCC/m}_{u\mathrm{SG}}=-\frac{4C^{\prime\prime}(|\vec{K}_{111}|)}{3}$ & $C^\mathrm{FCC/m}_{u\mathrm{SG}}=\frac{3(\hat{K}_i\cdot\hat{n})^2}{8}$\\
 & & & \\
- & - &$a^\mathrm{FCC/m}_{v\mathrm{SG}-L}=-C^{\prime\prime}(|\vec{G}_{200}|)$   &$C^\mathrm{FCC/m}_{v\mathrm{SG}-L}=\frac{(\hat{G}_i\cdot\hat{n})^2}{2}$\\
 & & & \\
- & - &$a^\mathrm{FCC/m}_{v\mathrm{SG}-T}=-\frac{2C^{\prime}(|\vec{G}_{200}|)}{|\vec{G}_{200}|}$   &$C^\mathrm{FCC/m}_{v\mathrm{SG}-T}=\frac{[1-(\hat{G}_i\cdot\hat{n})^2]}{4}$\\
\end{tabular}
\end{ruledtabular}
\label{tabA1}
\end{table*}

\begin{table*}[!htb]
\caption{List of symbols representing classifications of the density wave amplitude or the GL order parameters, the RLV subsets with respect to three BCC and three FCC CMIs. For each CMI orientation, the RLV are catogorized into subsets where $(\hat{K}_i\cdot\hat{n})^2$ or $(\hat{G}_i\cdot\hat{n})^2$ have the identical magnitude in each subset.}\begin{ruledtabular}
\begin{tabular}{lccccccccccccccccc}
&\multicolumn{2}{c}{BCC(100)}&\multicolumn{3}{c}{BCC(110)}&\multicolumn{2}{c}{BCC(111)}&\multicolumn{3}{c}{FCC(100)}&\multicolumn{4}{c}{FCC(110)}&\multicolumn{3}{c}{FCC(111)}\\
\hline
GL order parameters                                     
&\multicolumn{2}{c}{$u_a$ \ \  $u_b$}
&\multicolumn{3}{c}{$u_a$ \ \ $u_b$ \ \ $u_c$}
&\multicolumn{2}{c}{$u_a$ \ \ $u_b$}
&\multicolumn{3}{c}{$u_a$ \ \ $v_a$ \ \ $v_b$}
&\multicolumn{4}{c}{$u_a$ \ \ $u_b$ \ \ $v_a$ \ \ $v_b$}
&\multicolumn{3}{c}{$u_a$ \ \ $u_b$ \ \  $v_a$}\\
RLV subsets                                     
&\multicolumn{2}{c}{$\left\langle011\right\rangle$$\left\langle110\right\rangle$}
&\multicolumn{3}{c}{$\left\langle011\right\rangle$$\left\langle110\right\rangle$$\left\langle1\overline{1}0\right\rangle$}
&\multicolumn{2}{c}{$\left\langle110\right\rangle$$\left\langle1\overline{1}0\right\rangle$}
&\multicolumn{3}{c}{$\left\langle111\right\rangle$$\left\langle200\right\rangle$$\left\langle020\right\rangle$}
&\multicolumn{4}{c}{$\left\langle111\right\rangle$$\left\langle1\overline{1}1\right\rangle$$\left\langle200\right\rangle$$\left\langle002\right\rangle$}
&\multicolumn{3}{c}{$\left\langle111\right\rangle$$\left\langle1\overline{1}1\right\rangle$$\left\langle200\right\rangle$}\\
Number of $\vec{K}_i$ or $\vec{G}_i$ 
&\multicolumn{2}{c}{4 \ \ \ 8}&\multicolumn{3}{c}{\ 8 \ \ \ 2 \ \ \ 2}&\multicolumn{2}{c}{6 \ \ \ 6}&\multicolumn{3}{c}{8 \ \ \ 2 \  \ \ 4}&\multicolumn{4}{c}{4 \ \ \ \ 4 \ \ \ \ 4 \ \ \ \ 2}&\multicolumn{3}{c}{2 \ \ \ 6 \ \ \ 6}\\
$(\hat{K}_i\cdot\hat{n})^2$ or $(\hat{G}_i\cdot\hat{n})^2$
&\multicolumn{2}{c}{\ 0 \ \ 1/2}&\multicolumn{3}{c}{1/4 \ 1 \ \ \ 0}&\multicolumn{2}{c}{\ 0 \ \ 2/3}&\multicolumn{3}{c}{1/3 \  \ 1 \  \  0 \ }&\multicolumn{4}{c}{2/3  \ \ 0 \ \ \ 1/2 \ \ 0 \ }&\multicolumn{3}{c}{\ \ 1 \ \ 1/9  \ 1/3 \ }\\
\end{tabular}
\end{ruledtabular}
\label{tabA2}
\end{table*} 

The Kronecker delta functions in Eq.(\ref{eq:DeltaF-GL-BCC}) and Eq.(\ref{eq:DeltaF-GL-FCC}), ensure the free-energy functional contribution only under the condition for those RLVs form closed polygons, for example, $\vec{K}_{i}+\vec{K}_{j}+\cdots=0$. In Eq.(\ref{eq:DeltaF-GL-BCC}), the symmetries of the BCC RLVs result in nonvanishing cubic terms, which give rise to a free-energy barrier between the crystal and melt phases and guarantee a ﬁrst-order freezing transition. In Eq.(\ref{eq:DeltaF-GL-FCC}) for the case of FCC system, the cubic term would be absent if one only employs the principal RLV set, one has to include at least both the $\left\langle111\right\rangle$ principal RLVs and the $\left\langle200\right\rangle$ second set of RLVs to obtain a ﬁrst-order freezing transition.

The multiplicative coefficients ($a^\mathrm{BCC/m}$ and $a^\mathrm{FCC/m}$) of the quadratic terms in the GL model of the free-energy functionals, i.e., Eq.(\ref{eq:DeltaF-GL-BCC}) and Eq.(\ref{eq:DeltaF-GL-FCC}), can be obtained by comaring with above cDFT free-energy functionals, i.e., Eq.(\ref{eq:DeltaF-cdft-BCC}) and Eq.(\ref{eq:DeltaF-cdft-FCC}). The multiplicative coefficients of the higher power non-quadratic terms in Eq.(\ref{eq:DeltaF-GL-BCC}) and Eq.(\ref{eq:DeltaF-GL-FCC}) are derived by the constraints that (i) the equilibrium state of the crystal phase is a minimum of free energy, (ii) the crystal and the melt phases have equal free energy at $T=T_\mathrm{m}$.~\cite{Wu06,Wu16} The normalization are employed, i.e., normalizing the sum of the normalization coefficients ($C^\mathrm{BCC/m}$ and $C^\mathrm{FCC/m}$) to unity, for examples, $\sum_{\vec{K}_{i}}C^\mathrm{FCC/m}_{u\mathrm{SG}}=1$, or $\sum_{\vec{K}_{i},\vec{K}_{j}}C^\mathrm{BCC/m}_{uu}\delta_{0,\vec{K}_{i}+\vec{K}_{j}}=1$. More details on the employment of the normalization coefficients can be found in Refs.\onlinecite{Wu06,Wu16}.

For the FCC CMIs, the subscripts ``$T$'' and ``$L$'' in the symbols of the multiplicative and normalization coefficients of the square gradient terms for $v_i$, stand for the transverse and longitudinal components of $\hat{G}$.

All the multiplicative coefficients and the normalization coefficients employed in Eq.(\ref{eq:DeltaF-GL-BCC}) and Eq.(\ref{eq:DeltaF-GL-FCC}) are listed in Tab.~\ref{tabA1}. 

\label{sec:ap3}
\vspace{.3cm}
\noindent {\bf C. GL model formlism for specific CMI orientations}
\label{sec:ap3}
\vspace{.3cm}

Next, the GL formalism of the CMI free-energy functionals are reduced into forms for determining the CMI free energies $\gamma=\Delta F/A_{xy}$ of the specific CMI orientations. Note that, the density waves amplitudes of each RLVs set are equal in the bulk phases, while they are not equal in the interfacial region, leading to the CMI orientation-dependence on the magnitude of the CMI $\gamma$. Thus, one has to subdivide the GL order parameters into different catogories based on the direction cosine between the particular RLV and the CMI orientation $\hat{n}$, in order to derive the specific expression for the CMI $\gamma_{\hat{n}}$.~\cite{Wu06,Wu16} As shown in Tab.~\ref{tabA2}, base on the dot product of the CMI normal $\hat{n}$ and RLV $\hat{K}$ (or $\hat{G}$), GL order parameters in a given subset have the identical density wave amplitude across the CMI, are denoted by $u_a$, $u_b$, $\cdots$, (or $v_a$, $\cdots$). Subsets of density waves and their corresponding GL order parameters for the three low index CMI orientations of both BCC-melt and FCC-melt CMI systems, are summarized in Tab.~\ref{tabA2}. With explicitating the GL order parameters and the normalization coefficients of the square gradient terms, Eq.(\ref{eq:DeltaF-GL-BCC}) is reduced into three specific expressions of $\gamma^\mathrm{BCC/m}_{\hat{n}}$, i.e., Eq.(\ref{eq:DeltaF-GL-BCC100}) for $\gamma^\mathrm{BCC/m}_{100}$, Eq.(\ref{eq:DeltaF-GL-BCC110}) for $\gamma^\mathrm{BCC/m}_{110}$, and Eq.(\ref{eq:DeltaF-GL-BCC111}) for for $\gamma^\mathrm{BCC/m}_{111}$; Eq.(\ref{eq:DeltaF-GL-FCC}) is reduced into three specific expressions of $\gamma^\mathrm{FCC/m}_{\hat{n}}$, i.e., Eq.(\ref{eq:DeltaF-GL-FCC100}) for $\gamma^\mathrm{FCC/m}_{100}$, Eq.(\ref{eq:DeltaF-GL-FCC110}) for $\gamma^\mathrm{FCC/m}_{110}$, and Eq.(\ref{eq:DeltaF-GL-FCC111}) for $\gamma^\mathrm{FCC/m}_{111}$.

\begin{equation}
\begin{aligned}
\gamma^\mathrm{BCC/m}_{100} \approx &\frac{\rho_\mathrm{m}}{2\beta_\mathrm{m}} \int\mathrm{d}z  
\Bigg[a^\mathrm{BCC/m}_{uu}\left(\frac{1}{3}u_a^2+\frac{2}{3}u_b^2\right)-a^\mathrm{BCC/m}_{uuu}u_au_b^2
+a^\mathrm{BCC/m}_{uuuu}\left(\frac{1}{9}u_a^4+\frac{4}{9}u_a^2u_b^2+\frac{4}{9}u_b^4\right)
+a^\mathrm{BCC/m}_{u\mathrm{SG}}\left| \frac{\mathrm{d}u_b}{\mathrm{d}z}\right|^2\Bigg],
\label{eq:DeltaF-GL-BCC100}
\end{aligned}
\end{equation}

\begin{equation}
\begin{aligned}
\gamma^\mathrm{BCC/m}_{110} \approx &\frac{\rho_\mathrm{m}}{2\beta_\mathrm{m}}\int\mathrm{d}z 
\Bigg[a^\mathrm{BCC/m}_{uu}\left(\frac{4}{6}u_a^2+\frac{1}{6}u_b^2+\frac{1}{6}u_c^2\right)
-a^\mathrm{BCC/m}_{uuu}  \left( \frac{1}{2}u_a^2u_b+\frac{1}{2}u_a^2u_c \right) 
+a^\mathrm{BCC/m}_{uuuu} \bigg( \frac{12}{27}u_a^4+\frac{1}{27}u_b^4 \\ + &\frac{1}{27}u_c^4 + \frac{4}{27}u_a^2u_b^2
+\frac{4}{27}u_a^2u_c^2+\frac{1}{27}u_c^2u_b^2+\frac{4}{27}u_a^2u_bu_c  \bigg) 
+a^\mathrm{BCC/m}_{u\mathrm{SG}}\left( \frac{1}{2}\left| \frac{\mathrm{d}u_a}{\mathrm{d}z}\right|^2+ \frac{1}{2}\left| \frac{\mathrm{d}u_b}{\mathrm{d}z}\right|^2\right)\Bigg],
\label{eq:DeltaF-GL-BCC110}
\end{aligned}
\end{equation}

\begin{equation}
\begin{aligned}
\gamma^\mathrm{BCC/m}_{111} \approx &\frac{\rho_\mathrm{m}}{2\beta_\mathrm{m}}\int\mathrm{d}z  
\Bigg[a^\mathrm{BCC/m}_{uu}\left(\frac{1}{2}u_a^2+\frac{1}{2}u_b^2\right)-a^\mathrm{BCC/m}_{uuu}\left(\frac{3}{4}u_a^2u_b+\frac{1}{4}u_b^3\right)
+a^\mathrm{BCC/m}_{uuuu}\left(\frac{5}{9}u_a^2u_b^2+\frac{2}{9}u_a^4+\frac{2}{9}u_b^4\right)\\
+&a^\mathrm{BCC/m}_{u\mathrm{SG}}\left| \frac{\mathrm{d}u_a}{\mathrm{d}z}\right|^2\Bigg],
\label{eq:DeltaF-GL-BCC111}
\end{aligned}
\end{equation}

\begin{equation}
\begin{aligned}
\gamma^\mathrm{FCC/m}_{100} \approx &\frac{\rho_\mathrm{m}}{2\beta_\mathrm{m}}\int\mathrm{d}z  
\Bigg[a^\mathrm{FCC/m}_{uu}u_a^2-a^\mathrm{FCC/m}_{uuv}\left(\frac{1}{3}u_a^2v_a+\frac{2}{3}u_a^2v_b\right) 
+a^\mathrm{FCC/m}_{uuuu}u_a^4
+a^\mathrm{FCC/m}_{vv}\left(\frac{1}{3}v_a^2+\frac{2}{3}v_b^2\right)\\ 
+&a^\mathrm{FCC/m}_{uuvv}\left(\frac{1}{6}u_a^2v_a^2+\frac{3}{6}u_a^2v_b^2+\frac{2}{6}u_a^2v_av_b\right)
+a^\mathrm{FCC/m}_{u\mathrm{SG}}\left| \frac{\mathrm{d}u_a}{\mathrm{d}z}\right|^2
+a^\mathrm{FCC/m}_{v\mathrm{SG}-L}\left| \frac{\mathrm{d}v_a}{\mathrm{d}z}\right|^2+a^\mathrm{FCC/m}_{v\mathrm{SG}-T} \left| \frac{\mathrm{d}v_b}{\mathrm{d}z}\right|^2 \Bigg],
\label{eq:DeltaF-GL-FCC100}
\end{aligned}
\end{equation}

\begin{equation}
\begin{aligned}
\gamma^\mathrm{FCC/m}_{110} \approx &\frac{\rho_\mathrm{m}}{2\beta_\mathrm{m}}\int\mathrm{d}z  
\Bigg[a^\mathrm{FCC/m}_{uu}\left(\frac{1}{2}u_a^2+\frac{1}{2}u_b^2\right)-a^\mathrm{FCC/m}_{uuv}\left(\frac{1}{6}u_a^2v_b+\frac{1}{6}u_b^2v_b+\frac{4}{6}u_au_bv_a\right) 
+a^\mathrm{FCC/m}_{uuuu}\Bigg(\frac{1}{4}u_a^4+\frac{1}{4}u_b^4 \\ +& \frac{2}{4}u_a^2u_b^2\Bigg)+a^\mathrm{FCC/m}_{vv}\left(\frac{2}{3}v_a^2+\frac{1}{3}v_b^2\right) 
+a^\mathrm{FCC/m}_{uuvv}\left(\frac{3}{12}u_a^2v_a^2+\frac{1}{12}u_a^2v_b^2+\frac{3}{12}u_b^2v_a^2+\frac{1}{12}u_b^2v_b^2+\frac{4}{12}u_au_bv_av_b\right)\\
+&a^\mathrm{FCC/m}_{u\mathrm{SG}}\left| \frac{\mathrm{d}u_a}{\mathrm{d}z}\right|^2+a^\mathrm{FCC/m}_{v\mathrm{SG}-L}\left| \frac{\mathrm{d}v_a}{\mathrm{d}z}\right|^2
+a^\mathrm{FCC/m}_{v\mathrm{SG}-T}\left( \frac{1}{2}\left| \frac{\mathrm{d}v_a}{\mathrm{d}z}\right|^2+\frac{1}{2}\left| \frac{\mathrm{d}v_b}{\mathrm{d}z}\right|^2\right)\Bigg],
\label{eq:DeltaF-GL-FCC110}
\end{aligned}
\end{equation}

\begin{equation}
\begin{aligned}
\gamma^\mathrm{FCC/m}_{111} \approx &\frac{\rho_\mathrm{m}}{2\beta_\mathrm{m}}\int\mathrm{d}z  
\Bigg[a^\mathrm{FCC/m}_{uu}\left(\frac{1}{4}u_a^2+\frac{3}{4}u_b^2\right)-a^\mathrm{FCC/m}_{uuv}\left(\frac{1}{2}u_b^2v_a+\frac{1}{2}u_au_bv_a\right) 
+a^\mathrm{FCC/m}_{uuuu} \Bigg( \frac{1}{12}u_a^4+\frac{6}{12}u_b^4+ \frac{3}{12}u_a^2u_b^2 \\+ &\frac{2}{12}u_au_b^3 \Bigg) +a^\mathrm{FCC/m}_{vv}v_a^2 
+a^\mathrm{FCC/m}_{uuvv}\left(\frac{1}{8}u_a^2v_a^2+\frac{5}{8}u_b^2v_a^2+\frac{2}{8}u_au_bv_a^2\right)
+a^\mathrm{FCC/m}_{u\mathrm{SG}}\left(\frac{3}{4}\left| \frac{\mathrm{d}u_a}{\mathrm{d}z}\right|^2+\frac{1}{4}\left| \frac{\mathrm{d}u_b}{\mathrm{d}z}\right|^2\right)\\
+&a^\mathrm{FCC/m}_{v\mathrm{SG}-L}\left| \frac{\mathrm{d}v_a}{\mathrm{d}z}\right|^2+a^\mathrm{FCC/m}_{v\mathrm{SG}-T} \left| \frac{\mathrm{d}v_a}{\mathrm{d}z}\right|^2 \Bigg],
\label{eq:DeltaF-GL-FCC111}
\end{aligned}
\end{equation}

The contribution components of the $\gamma$ are inspected, include the component $\gamma_\mathrm{nonSG}$ arises from the non-square gradient term and the other component(s) arise from the square gradient terms ($\gamma_{u\mathrm{SG}}$ and/or $\gamma_{v\mathrm{SG}}$) of the GL OPPs due to the principal and/or the second set of RLVs. For example, for the FCC(100) CMI, the Eq.(\ref{eq:DeltaF-GL-FCC100}) can be rewritten as the following:

\begin{subequations}
\begin{align}
\gamma^\mathrm{FCC/m}_{100} = \gamma^{\mathrm{FCC/m}}_{100,\mathrm{nonSG}} +\gamma^{\mathrm{FCC/m}}_{100,{u\mathrm{SG}}} +\gamma^{\mathrm{FCC/m}}_{100,{v\mathrm{SG}}}
\end{align}

\begin{align}
\begin{aligned}
\gamma^{\mathrm{FCC/m}}_{100,\mathrm{nonSG}} \approx &\frac{\rho_\mathrm{m}}{2\beta_\mathrm{m}}\int\mathrm{d}z  
\Bigg[a^\mathrm{FCC/m}_{uu}u_a^2-a^\mathrm{FCC/m}_{uuv}\left(\frac{1}{3}u_a^2v_a+\frac{2}{3}u_a^2v_b\right) 
+a^\mathrm{FCC/m}_{uuuu}u_a^4
+a^\mathrm{FCC/m}_{vv}\left(\frac{1}{3}v_a^2+\frac{2}{3}v_b^2\right)\\ 
+&a^\mathrm{FCC/m}_{uuvv}\left(\frac{1}{6}u_a^2v_a^2+\frac{3}{6}u_a^2v_b^2+\frac{2}{6}u_a^2v_av_b\right) \Bigg]
\end{aligned}
\end{align}

\begin{align}
\gamma^{\mathrm{FCC/m}}_{100,{u\mathrm{SG}}} \approx \frac{\rho_\mathrm{m}}{2\beta_\mathrm{m}}\int\mathrm{d}z \ a^\mathrm{FCC/m}_{u\mathrm{SG}}\left| \frac{\mathrm{d}u_a}{\mathrm{d}z}\right|^2
\end{align}

\begin{align}
\gamma^{\mathrm{FCC/m}}_{100,{v\mathrm{SG}}} \approx  \frac{\rho_\mathrm{m}}{2\beta_\mathrm{m}}\int\mathrm{d}z \Bigg( a^\mathrm{FCC/m}_{v\mathrm{SG}-L}\left| \frac{\mathrm{d}v_a}{\mathrm{d}z}\right|^2+a^\mathrm{FCC/m}_{v\mathrm{SG}-T} \left| \frac{\mathrm{d}v_b}{\mathrm{d}z}\right|^2 \Bigg) 
\end{align}

\end{subequations}

\newpage

\vspace{1.5cm}
\noindent {\bf II. Analytically minimization of the free energy functionals}
\label{sec:ap3}
\vspace{.3cm}

Consider a Functional $\mathcal{F}[u(z)]$ of the form
\begin{equation}
	\mathcal{F}[u(z)]=\int_{a}^{b}f\left[z,u(z),\frac{\mathrm{d}u(z)}{\mathrm{d}z}\right]\mathrm{d}z=\int_{a}^{b}f\left[z,u(z),u'(z)\right]\mathrm{d}z,
\label{eqFu}
\end{equation}
where $f$ is the integrand function and the Euler-Lagrange equation
\begin{equation}
	\left(\frac{\partial f}{\partial u}\right)-\frac{\mathrm{d}}{\mathrm{d}z}\left(\frac{\partial f}{\partial u'}\right)=0.
\label{eqEL}
\end{equation}

With the analytical expressions of the interfacial free energies, Eqs.(A12-A17), in the Appendix of the main text. We can generate coupled Euler-Lagrange (cE-L) equations for each of the GL OPPs, $u_i(z)$ and $v_i(z)$.

For BCC(100) CMI,

\begin{equation}
\begin{aligned}
	&\left(\frac{\partial f}{\partial u_a}\right)-\frac{\mathrm{d}}{\mathrm{d}z}\left(\frac{\partial f}{\partial u_a'}\right)=0,\\
	&\left(\frac{\partial f}{\partial u_b}\right)-\frac{\mathrm{d}}{\mathrm{d}z}\left(\frac{\partial f}{\partial u_b'}\right)=0.
\end{aligned}
\end{equation}

By computing the partial derivatives for each field, the full cE-L equations for each GL OPPs are as follows,

\begin{equation}
\begin{aligned}
&a^\mathrm{BCC/m}_{uu}\left(\frac{2}{3}u_a(z)\right)-a^\mathrm{BCC/m}_{uuu}u_b^2(z)+a^\mathrm{BCC/m}_{uuuu}\left(\frac{4}{9}u_a^3(z)+\frac{8}{9}u_a(z)u_b^2(z)\right)=0,\\
\\
&a^\mathrm{BCC/m}_{uu}\left(\frac{4}{3}u_b(z)\right)-2a^\mathrm{BCC/m}_{uuu}u_a(z)u_b(z)+a^\mathrm{BCC/m}_{uuuu}\left(\frac{8}{9}u_a^2(z)u_b(z)+\frac{16}{9}u_b^3(z)\right)-2a^\mathrm{BCC/m}_{u\mathrm{SG}}u_b''(z)=0.
\end{aligned}
\end{equation}

For BCC(110) CMI,
\begin{equation}
\begin{aligned}
	\left(\frac{\partial f}{\partial u_a}\right)-\frac{\mathrm{d}}{\mathrm{d}z}\left(\frac{\partial f}{\partial u_a'}\right)=0,\\
	\left(\frac{\partial f}{\partial u_b}\right)-\frac{\mathrm{d}}{\mathrm{d}z}\left(\frac{\partial f}{\partial u_b'}\right)=0,\\
	\left(\frac{\partial f}{\partial u_c}\right)-\frac{\mathrm{d}}{\mathrm{d}z}\left(\frac{\partial f}{\partial u_c'}\right)=0,
\end{aligned}
\end{equation}

The full cE-L equations for each GL OPPs are as follows,

\begin{equation}
\begin{aligned}
&a^\mathrm{BCC/m}_{uu}\left(\frac{4}{3}u_a(z)\right)-a^\mathrm{BCC/m}_{uuu}\left[ u_a(z)u_b(z)+u_a(z)u_c(z) \right]+\\
&a^\mathrm{BCC/m}_{uuuu}\left(\frac{48}{27}u_a^3(z)+\frac{8}{27}u_a(z)u_b^2(z)+\frac{8}{27}u_a(z)u_c^2(z)+\frac{8}{27}u_a(z)u_b(z)u_c(z)\right)-a^\mathrm{BCC/m}_{u\mathrm{SG}}u_a''(z)=0,\\
\\
&a^\mathrm{BCC/m}_{uu}\left(\frac{1}{3}u_b(z)\right)-a^\mathrm{BCC/m}_{uuu}\left(\frac{1}{2}u_a^2(z)\right)+\\
&a^\mathrm{BCC/m}_{uuuu}\left(\frac{4}{27}u_b^3(z)+\frac{8}{27}u_a^2(z)u_b(z)+\frac{2}{27}u_c^2(z)u_b(z)+\frac{4}{27}u_a^2(z)u_c(z)\right)-a^\mathrm{BCC/m}_{u\mathrm{SG}}u_b''(z)=0,\\
\\
&a^\mathrm{BCC/m}_{uu}\left(\frac{1}{3}u_c(z)\right)-a^\mathrm{BCC/m}_{uuu}\left(\frac{1}{2}u_a^2(z)\right)+\\
&a^\mathrm{BCC/m}_{uuuu}\left(\frac{4}{27}u_c^3(z)+\frac{8}{27}u_a^2(z)u_c(z)+\frac{2}{27}u_c(z)u_b^2(z)+\frac{4}{27}u_a^2(z)u_b(z)\right)=0.
\end{aligned}
\end{equation}

For BCC(111) CMI,
\begin{equation}
\begin{aligned}
	&\left(\frac{\partial f}{\partial u_a}\right)-\frac{\mathrm{d}}{\mathrm{d}z}\left(\frac{\partial f}{\partial u_a'}\right)=0,\\
	&\left(\frac{\partial f}{\partial u_b}\right)-\frac{\mathrm{d}}{\mathrm{d}z}\left(\frac{\partial f}{\partial u_b'}\right)=0
\end{aligned}
\end{equation}

The full cE-L equations for each GL OPPs are as follows,

\begin{equation}
\begin{aligned}
&a^\mathrm{BCC/m}_{uu}u_a(z)-a^\mathrm{BCC/m}_{uuu}\left(\frac{3}{2}u_a(z)u_b(z)\right)+a^\mathrm{BCC/m}_{uuuu}\left(\frac{10}{9}u_a(z)u_b^2(z)+\frac{8}{9}u_a^3(z)\right)-2a^\mathrm{BCC/m}_{u\mathrm{SG}}u_a''(z)=0,\\
\\
&a^\mathrm{BCC/m}_{uu}u_b(z)-a^\mathrm{BCC/m}_{uuu}\left(\frac{3}{4}u_a^2(z)+\frac{3}{4}u_b^2(z)\right)+a^\mathrm{BCC/m}_{uuuu}\left(\frac{10}{9}u_a^2(z)u_b(z)+\frac{8}{9}u_b^3(z)\right)=0.
\end{aligned}
\end{equation}

\newpage

For FCC(100) CMI,
\begin{equation}
\begin{aligned}
	&\left(\frac{\partial f}{\partial u_a}\right)-\frac{\mathrm{d}}{\mathrm{d}z}\left(\frac{\partial f}{\partial u_a'}\right)=0,\\
	&\left(\frac{\partial f}{\partial v_a}\right)-\frac{\mathrm{d}}{\mathrm{d}z}\left(\frac{\partial f}{\partial v_a'}\right)=0,\\
	&\left(\frac{\partial f}{\partial v_b}\right)-\frac{\mathrm{d}}{\mathrm{d}z}\left(\frac{\partial f}{\partial v_b'}\right)=0,
\end{aligned}
\end{equation}

The full cE-L equations for each GL OPPs are as follows,

\begin{equation}
\begin{aligned}
	&2a^\mathrm{FCC/m}_{uu}u_a(z)-a^\mathrm{FCC/m}_{uuv}\left(\frac{2}{3}u_a(z)v_a(z)+\frac{4}{3}u_a(z)v_b(z)\right)+4a^\mathrm{FCC/m}_{uuuu}u_a^3(z)+\\
	&a^\mathrm{FCC/m}_{uuvv}\left(\frac{1}{3}u_a(z)v_a^2(z)+u_a(z)v_b^2(z)+\frac{2}{3}u_a(z)v_a(z)v_b(z)\right)-2a^\mathrm{FCC/m}_{u\mathrm{SG}}u_a''(z)=0,\\
	\\
	&-a^\mathrm{FCC/m}_{uuv}\left(\frac{1}{3}u_a^2(z)\right)+a^\mathrm{FCC/m}_{vv}\left(\frac{2}{3}v_a(z)\right)+\\
	&a^\mathrm{FCC/m}_{uuvv}\left(\frac{1}{3}u_a^2(z)v_a(z)+\frac{1}{3}u_a^2(z)v_b(z)\right)-2a^\mathrm{FCC/m}_{v\mathrm{SG}-L}v_a''(z)=0,\\
	\\
	&-a^\mathrm{FCC/m}_{uuv}\left(\frac{2}{3}u_a^2(z)\right)+a^\mathrm{FCC/m}_{vv}\left(\frac{4}{3}v_b(z)\right)+\\
	&a^\mathrm{FCC/m}_{uuvv}\left(u_a^2(z)v_b(z)+\frac{1}{3}u_a^2(z)v_a(z)\right)-2a^\mathrm{FCC/m}_{v\mathrm{SG}-T}v_b''(z)=0.
\end{aligned}
\end{equation}

\newpage
For FCC(110) CMI,
\begin{equation}
\begin{aligned}
	&\left(\frac{\partial f}{\partial u_a}\right)-\frac{\mathrm{d}}{\mathrm{d}z}\left(\frac{\partial f}{\partial u_a'}\right)=0,\\
	&\left(\frac{\partial f}{\partial u_b}\right)-\frac{\mathrm{d}}{\mathrm{d}z}\left(\frac{\partial f}{\partial u_b'}\right)=0,\\
	&\left(\frac{\partial f}{\partial v_a}\right)-\frac{\mathrm{d}}{\mathrm{d}z}\left(\frac{\partial f}{\partial v_a'}\right)=0,\\
	&\left(\frac{\partial f}{\partial v_b}\right)-\frac{\mathrm{d}}{\mathrm{d}z}\left(\frac{\partial f}{\partial v_b'}\right)=0,
\end{aligned}
\end{equation}

The full cE-L equations for each GL OPPs are as follows,

\begin{equation}
\begin{aligned}
&a^\mathrm{FCC/m}_{uu}u_a(z)-a^\mathrm{FCC/m}_{uuv}\left(\frac{1}{3}u_a(z)v_b(z)+\frac{2}{3}u_b(z)v_a(z)\right)+a^\mathrm{FCC/m}_{uuuu}\left(u_a^3(z)+u_a(z)u_b^2(z)\right)+\\
&a^\mathrm{FCC/m}_{uuvv}\left(\frac{1}{2}u_a(z)v_a^2(z)+\frac{1}{6}u_a(z)v_b^2(z)+\frac{1}{3}u_b(z)v_a(z)v_b(z)\right)-2a^\mathrm{FCC/m}_{u\mathrm{SG}}u_a''(z)=0,\\
\\
&a^\mathrm{FCC/m}_{uu}u_b(z)-a^\mathrm{FCC/m}_{uuv}\left(\frac{1}{3}u_b(z)v_b(z)+\frac{2}{3}u_a(z)v_a(z)\right)+a^\mathrm{FCC/m}_{uuuu}\left(u_b^3(z)+u_a^2(z)u_b(z)\right)+\\
&a^\mathrm{FCC/m}_{uuvv}\left(\frac{1}{2}u_b(z)v_a^2(z)+\frac{1}{6}u_b(z)v_b^2(z)+\frac{1}{3}u_a(z)v_a(z)v_b(z)\right)=0,\\
\\
&-a^\mathrm{FCC/m}_{uuv}\left(\frac{2}{3}u_a(z)u_b(z)\right)+a^\mathrm{FCC/m}_{vv}\left(\frac{4}{3}v_a(z)\right)+\\
&a^\mathrm{FCC/m}_{uuvv}\left(\frac{1}{2}u_a^2(z)v_a(z)+\frac{1}{2}u_b^2(z)v_a(z)+\frac{1}{3}u_a(z)u_b(z)v_b(z)\right)-\left(2a^\mathrm{FCC/m}_{v\mathrm{SG}-L}+a^\mathrm{FCC/m}_{v\mathrm{SG}-T}\right)v_a''(z)=0,\\
\\
&-a^\mathrm{FCC/m}_{uuv}\left(\frac{1}{6}u_a^2(z)+\frac{1}{6}u_b^2(z)\right)+a^\mathrm{FCC/m}_{vv}\left(\frac{2}{3}v_b(z)\right)+\\
&a^\mathrm{FCC/m}_{uuvv}\left(\frac{1}{6}u_a^2(z)v_b(z)+\frac{1}{6}u_b^2(z)v_b(z)+\frac{1}{3}u_a(z)u_b(z)v_a(z)\right)-a^\mathrm{FCC/m}_{v\mathrm{SG}-T}v_b''(z)=0.\\
\end{aligned}
\end{equation}

\newpage
For FCC(111) CMI,
\begin{equation}
\begin{aligned}
	&\left(\frac{\partial f}{\partial u_a}\right)-\frac{\mathrm{d}}{\mathrm{d}z}\left(\frac{\partial f}{\partial u_a'}\right)=0,\\
	&\left(\frac{\partial f}{\partial u_b}\right)-\frac{\mathrm{d}}{\mathrm{d}z}\left(\frac{\partial f}{\partial u_b'}\right)=0,\\
	&\left(\frac{\partial f}{\partial v_a}\right)-\frac{\mathrm{d}}{\mathrm{d}z}\left(\frac{\partial f}{\partial v_a'}\right)=0,
\end{aligned}
\end{equation}

The full cE-L equations for each GL OPPs are as follows,

\begin{equation}
\begin{aligned}
&a^\mathrm{FCC/m}_{uu}\left(\frac{1}{2}u_a(z)\right)-a^\mathrm{FCC/m}_{uuv}\left(\frac{1}{2}u_b(z)v_a(z)\right)+a^\mathrm{FCC/m}_{uuuu}\left(\frac{1}{3}u_a^3(z)+\frac{1}{2}u_a(z)u_b^2(z)+\frac{1}{6}u_b^3(z)\right)+\\
&a^\mathrm{FCC/m}_{uuvv}\left(\frac{1}{4}u_a(z)v_a^2(z)+\frac{1}{4}u_b(z)v_a^2(z)\right)-\frac{3}{2}a^\mathrm{FCC/m}_{u\mathrm{SG}}u_a''(z)=0,\\
\\
&a^\mathrm{FCC/m}_{uu}\left(\frac{3}{2}u_b(z)\right)-a^\mathrm{FCC/m}_{uuv}\left(u_b(z)v_a(z)+\frac{1}{2}u_a(z)v_a(z)\right)+\\
&a^\mathrm{FCC/m}_{uuuu}\left(2u^3_b(z)+\frac{1}{2}u_a^2(z)u_b(z)+\frac{1}{2}u_a(z)u_b^2(z)\right)+\\
&a^\mathrm{FCC/m}_{uuvv}\left(\frac{5}{4}u_b(z)v_a^2(z)+\frac{1}{4}u_a(z)v_a^2(z)\right)-\frac{1}{2}a^\mathrm{FCC/m}_{u\mathrm{SG}}u_b''(z)=0,\\
\\
&-a^\mathrm{FCC/m}_{uuv}\left(\frac{1}{2}u_b^2(z)+\frac{1}{2}u_a(z)u_b(z)\right)+2a^\mathrm{FCC/m}_{vv}v_a(z)+\\
&a^\mathrm{FCC/m}_{uuvv}\left(\frac{1}{4}u_a^2(z)v_a(z)+\frac{5}{4}u_b^2(z)v_a(z)+\frac{1}{2}u_a(z)u_b(z)v_a(z)\right)-2\left(a^\mathrm{FCC/m}_{v\mathrm{SG}-L}+a^\mathrm{FCC/m}_{v\mathrm{SG}-T}\right)v_a''(z)=0.
\end{aligned}
\end{equation}

Above cE-L  equations are solved numerically subject to the boundary conditions that these GL OPPs vanish in bulk melt phase and retain ﬁxed values in bulk crystals, i.e., $u_\mathrm{s}$ (and $v_\mathrm{s}$).

\newpage

\vspace{1.5cm}
\noindent {\bf III. Supplemental table}
\label{sec:ap1}
\vspace{.3cm}

\begin{table*}[!htb]
\caption{Summary of the geometries and sizes of the simulation cells for the six CMIs investigated.}
\begin{ruledtabular}
\begin{tabular}{c c c }
~~Orientation~~&~~~Size~~&~~$N$~~\\
\hline
BCC(100)& 124 $\sigma$ $\times$ 124 $\sigma$ $\times$ 134 $\sigma$ 
&47320
\\
BCC(110)& 124 $\sigma$ $\times$ 121 $\sigma$ $\times$ 135 $\sigma$ 
&46566
\\
BCC(111)& 121 $\sigma$ $\times$ 116 $\sigma$ $\times$ 141 $\sigma$ 
&45900
\\
FCC(100)& 119$\sigma$ $\times$ 119$\sigma$ $\times$ 132$\sigma$ 
&44000
\\
FCC(110)& 119$\sigma$ $\times$ 118$\sigma$ $\times$ 133$\sigma$ 
&43680
\\
FCC(111)& 118$\sigma$ $\times$ 117$\sigma$ $\times$ 136$\sigma$ 
&43680
\end{tabular}
\end{ruledtabular}
\label{tabs1}
\end{table*}

\newpage

\vspace{1.5cm}
\noindent {\bf IV. Supplemental figures}
\label{sec:ap2}
\vspace{.3cm}

 \begin{figure}[!htb]
\centering
 \includegraphics[width=0.75\textwidth]{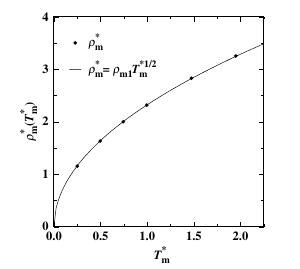}
 \caption{The densities of the uniform bulk melt phases $\rho^*_\mathrm{m}$ (at the crystal-melt coexistence temperature and pressure) follow well the corresponding power-law scaling.}
 \label{figA1}
 \end{figure} 

 \begin{figure}[!htb]
\centering
 \includegraphics[width=0.95\textwidth]{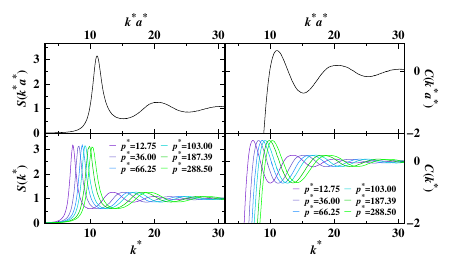}
 \caption{Structure factors (left), and direct correlation functions (right), as functions of the dimensionless wavenumber $k^*a^*$ (top) and as functions of the original wavenumbers $k^*$ (bottom).}
 \label{figA2}
 \end{figure} 
 
 \begin{figure}[!htb]
\centering
 \includegraphics[width=0.95\textwidth]{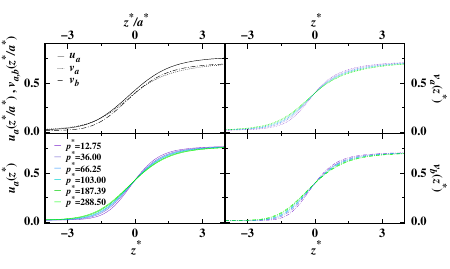}
 \caption{GL order parameter (amplitudes of density waves, $u$, $v_a$, $v_b$) profiles, as the function of the dimensionless distance ($z^*/a^*$) relative to the CMI position $(\xi^*/a^*)(t^*)$, across equilibrium SS FCC(100) CMI. $z^*/a^*>0$ for melt and $z^*/a^*<0$ for crystal (left top panel). The rest three panels present the three individial GL order parameter profiles, as functions of the original distance $z^*$ relative to the CMI position, under six different $p^*$ along the coexistence boundary.}
 \label{figA3}
 \end{figure} 
 
\begin{figure}[!htb]
\centering
 \includegraphics[width=0.75\textwidth]{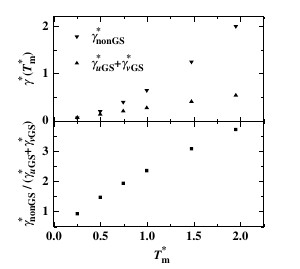}
 \caption{Top panel, The GL model predictions of 
the contribution components of the $\gamma^\mathrm{FCC/m*}_{100}$, $\gamma^*_\mathrm{nonSG}$  arises from the non-square gradient term and the other component(s) arise from the square gradient terms (the sum of $\gamma^*_{u\mathrm{SG}}$ and $\gamma^*_{v\mathrm{SG}}$) under six different $p^*$ along the coexistence boundary. Bottom panel, the relative contributions of non-gradient and square gradient terms.}
 \label{figA4}
 \end{figure} 

\clearpage

\bibliography{SM}
